\documentclass[pdflatex,sn-aps]{sn-jnl}

\usepackage{graphicx}
\usepackage{dcolumn}
\usepackage{hyperref}
\usepackage{comment}
\usepackage{xcolor}
\usepackage{amsmath}
\usepackage{mathrsfs}
\usepackage{mathtools}

\usepackage{caption}
\usepackage{subcaption}
\begin{document}

%%%%%%%%%%%%%%%%%%%%%%%%%%%%%%%%%%%%%%%%%%%%%%%%%%%%%%%%%
%                       ~Title~                         %
%%%%%%%%%%%%%%%%%%%%%%%%%%%%%%%%%%%%%%%%%%%%%%%%%%%%%%%%%

\title{System design and realisation towards optimising secure key bits in free space QKD}

\author*[1]{\fnm{Pooja} \sur{Chandravanshi}}\email{pooja@prl.res.in}

\author*[1]{\fnm{Jayanth}\sur{Ramakrishnan}}\email{jayanthrg98@gmail.com}

\author[1,3]{\fnm{Tanya} \sur{Sharma}}

\author*[2]{\fnm{Ayan} \sur{Biswas}}\email{b.ayan13092@gmail.com}

\author*[1]{\fnm{Ravindra P.} \sur{Singh}}\email{rpkpsingh@gmail.com}

\affil[1]{\orgdiv{Quantum Science and Technology Laboratory}, \orgname{Physical Research Laboratory, Department of Space,Govt. of India}, \orgaddress{\city{Ahmedabad}, \postcode{380015}, \state{Gujarat}, 
\country{India}}}

\affil[2]{\orgdiv{Centre for Development of Telematics}, \orgname{Telecom Technology Centre of Govt. of India}, \orgaddress{\street{} \city{Delhi}, \postcode{110030}, 
\country{India}}}

\affil[3]{\orgdiv{Telecom Paris,
} \orgname{Institut Polytechnique de Paris}, \orgaddress{\street{} \city{19 Place Marguerite Perey}, \postcode{91120 Palaiseau}, 
\country{France}}}


\date{\today}

\abstract{Quantum Key Distribution (QKD) is rapidly transitioning from cutting-edge laboratory research to real-world deployment in established communication networks. Although QKD promises future-proof security, practical challenges still exist due to imperfections in physical devices. Many protocols offer strong security guarantees, but their implementation can be complex and difficult. To bridge this gap, we present a practical and systematic framework for implementing QKD, focused on the BB84 protocol but designed with broader applicability in mind. The article includes key concepts for device calibration, synchronisation, optical alignment, and key post-processing. We outline a simple algorithm for key sifting that is easily implementable in hardware. Our results highlight the importance of selecting the temporal window to optimise both the key rate and the quantum bit error rate (QBER). In addition, we show that random sampling of the sifted key bits for error estimation yields more reliable results than sequential sampling. We also integrate the Entrapped Pulse Coincidence Detection (EPCD) protocol to boost key generation rates, further enhancing performance. Although our work focuses on BB84, the techniques and practices outlined are general enough to support a wide range of QKD protocols. This makes our framework a valuable tool for both research and real-world deployment of secure quantum communication systems.}

\maketitle

\section{\label{sec:Introduction}Introduction}

In the world of expanding digital connectivity, the security of sensitive information is more crucial than ever. While effective, traditional cryptographic systems face ongoing threats from advances in quantum computing and algorithms. To address these vulnerabilities, Quantum Key Distribution (QKD) \cite{BB84} emerges as a groundbreaking solution, harnessing the principles of quantum mechanics to create unhackable communication channels.
QKD uses principles of quantum mechanics, such as the uncertainty principle and the no-cloning theorem, to ensure the security of data exchanged between parties. Due to the fundamental quantum mechanical principle that measurement disturbs the state, QKD uniquely offers the inherent capability to detect any eavesdropping attempt. Simply stated, QKD enables two parties, commonly called Alice and Bob, to securely share a secret key that can be used for subsequent encryption and secure communication.
QKD protocols are broadly classified based on the source and the type of variables used. Depending on the source, QKD can be categorised as either prepare and measure (P$\&$M) based or entanglement based. Additionally, based on the type of variables used, QKD protocols are divided into Discrete Variable (DV) and Continuous Variable (CV) protocols. DV-QKD encodes quantum information in discrete states, such as photon polarisation, and typically uses single-photon detectors for measurement. In contrast, CV-QKD encodes information in continuous properties of the electromagnetic field, such as amplitude and phase, and relies on detection techniques like homodyne or heterodyne measurement. A new secure communication paradigm, Quantum Secure Direct Communication (QSDC) \cite{pan2025simultaneous,pan2023free}, provides a near-instantaneous secure communication solution by encoding the actual messages on the quantum bits, rather than employing secret keys, as in QKD.

The reliable deployment of any prepare and measure  QKD protocol hinges on three core components, source, channel, and receiver, each with stringent requirements. At the source, laser diodes driven by high-speed electronics are typically used to prepare states \cite{laserdriver,zhu2022laser}. Crucial to this stage is the use of robust randomisation techniques to ensure truly random state preparation \cite{martin2015quantum, zhou2019practical}, mitigating any exploitable patterns. For the quantum channel, which connects the communicating parties, key physical parameters such as channel length, transmittance, and optical losses must be analysed, particularly in real-world implementation, where these factors significantly affect the fidelity of quantum state transmission \cite{dynes2016quantum,cai2019experimental,chou2023satellite}. The receiver component demands precise calibration of detector-related parameters, including dead time, dark count rate, and background noise \cite{jain2011device,fei2018quantum}. Accurate and precise characterisation at each of these stages is vital to maintain the efficiency and trustworthiness of the system. Beyond the physical layers, clock synchronisation between transmitter and receiver is indispensable for secure key generation \cite{miller2023time, krause2025clock}. In practical systems, the Data Acquisition System (DAQ) plays a central role in real-time capture, processing, and handling of quantum and classical data throughout the QKD process \cite{zhang2012real,shen2013fpga,yang2020high,stanco2022versatile}. End-to-end secure key generation at both Alice’s and Bob’s nodes follows a well-defined chain: quantum state preparation, transmission, detection, time stamping, sifting, parameter estimation, error correction, and privacy amplification.

Since its theoretical inception in 1984 and initial laboratory demonstrations in 1989 \cite{10.1145/74074.74087}, QKD has evolved dramatically, culminating in large-scale intercontinental implementations (e.g., the Micius satellite-based experiment). Modern QKD systems now span both fiber and free-space platforms \cite{Lucamarini2018, alleaume2014using, diamanti2016practical, kim2008implementation, schmitt2007experimental, namazi2017free} moving towards compact chip based systems \cite{sibson2017chip,huang2023chip}. Despite this progress, comprehensive practical insights spanning source and detector design, channel characterisation, and post-processing workflows remain relatively underrepresented in the literature. This gap underscores the need for continued investigation into fully integrated, real-world QKD systems.
Our work aims to bridge this gap by providing a detailed overview of QKD implementation in both laboratory and field settings for free-space QKD.
This work provides a clear and comprehensive guide to QKD implementation, with a focus on the BB84 protocol. Before implementing QKD, we meticulously characterise the source, channel, and detectors. All subsequent post-processing steps are integrated into the Data Acquisition (DAQ) system to facilitate real-time key generation.
In addition, with BB84, we have extended our implementation to incorporate 
EPCD protocol.

The structure of this article is as follows: Section \ref{sec:Theory} outlines the theoretical framework of prepare and measure QKD, covering BB84 protocol execution, secure key rate under practical conditions and theoretical background to EPCD protocol. Section \ref{sec:pre} details the experimental characterisation of the source and detector with respect to side-channel vulnerabilities. It also describes channel characterisation and time synchronisation between the transmitter
and receiver. Section \ref{Sec: practical} outlines the transmitter, receiver, and channel components, including the laser diode driver module, RNG for enabling laser diodes, polarisation encoding and decoding optics, launching and collection optics. It also describes interfacing and automation with channel arrangement. Section \ref{Sec:field} presents the field implementation of the protocol, which includes real-time protocol execution and post-processing in detail. Results are discussed in Section \ref{sec:RnD}, and conclusions are drawn in Section \ref{sec:conc}.
%%%%%%%%%%%%%%%%%%%%%%%%%%%%%%%%%%%%%%%%%%%%%%%%%%%%%%%%%
%             ~TheoreticalBackground~                  %
%%%%%%%%%%%%%%%%%%%%%%%%%%%%%%%%%%%%%%%%%%%%%%%%%%%%%%%%%
\section{\label{sec:Theory} Theoretical Background}
\subsection{QKD Protocol: Prepare and Measure}

\begin{figure}[h]
    \centering \includegraphics[width=10cm]{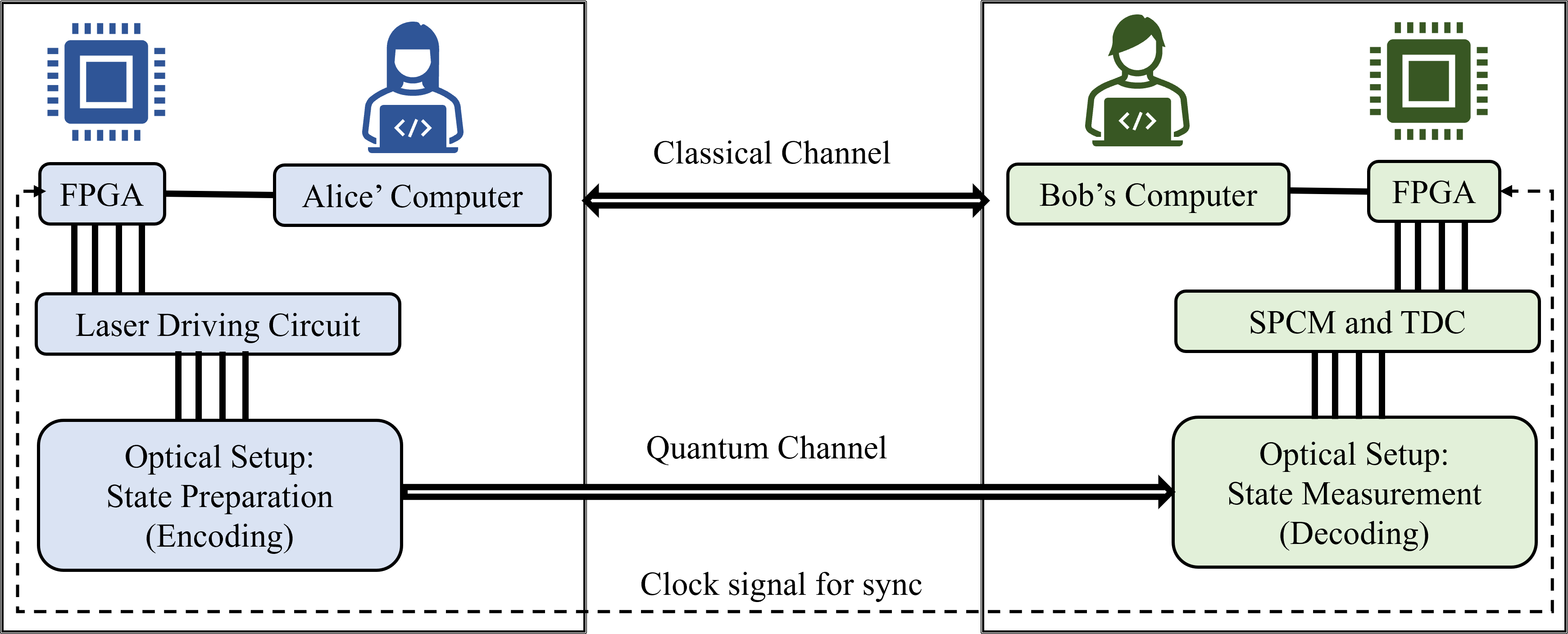}\\
     
    \caption{Basic block diagram of the prepare and measure QKD protocol. FPGA: Field Programmable Gate Array, SPCM: Single Photon Counting Module, TDC: Time to Digital Converter. Quantum channel and classical channel can be free space or optical fiber. }
    \label{fig: block}
\end{figure}

In P$\&$M QKD protocols, Alice prepares quantum states (qubits) in randomly chosen bases and sends them to Bob, who independently measures them using randomly selected bases. By publicly comparing their basis choices, Alice and Bob can extract a raw key from the matched events. When their basis choices align, Bob’s measurements and Alice's choice of bits are correlated, hence yielding valid key bits. Otherwise, the results are discarded. Any attempt at eavesdropping disturbs the quantum states, introducing detectable errors that reveal the presence of the intruder.
The final secure key is then distilled from the raw key
through classical post-processing procedures \cite{luo2024overview}. The primary components of QKD post-processing  include:
\begin{enumerate}
    \item Sifting: In this process, Alice and Bob retain only those bits where they use the same basis and discard the rest.
    \item Parameter Estimation: This involves estimating various parameters that affect the security and efficiency of the key generation process, including the QBER and potential information leakage to eavesdroppers.
    \item Error Correction: This step rectifies discrepancies between Alice's and Bob's keys due to noise and errors in transmission. 
    \item Privacy Amplification: After error correction, privacy amplification reduces any information that an eavesdropper may have gained during transmission, ensuring that the final keys are secure.
\end{enumerate}
Fig.\ref{fig: block} is the basic block diagram of the P$\&$M QKD protocol, depicting all the important components. 
In this paper, we discuss the BB84 QKD protocol using the polarisation degree of freedom. However, the principles outlined here can be extended to other degrees of freedom as well.

\subsubsection{\textbf{ BB84 Protocol}}

\begin{figure}[h]
    \centering
     \includegraphics[width=10cm]{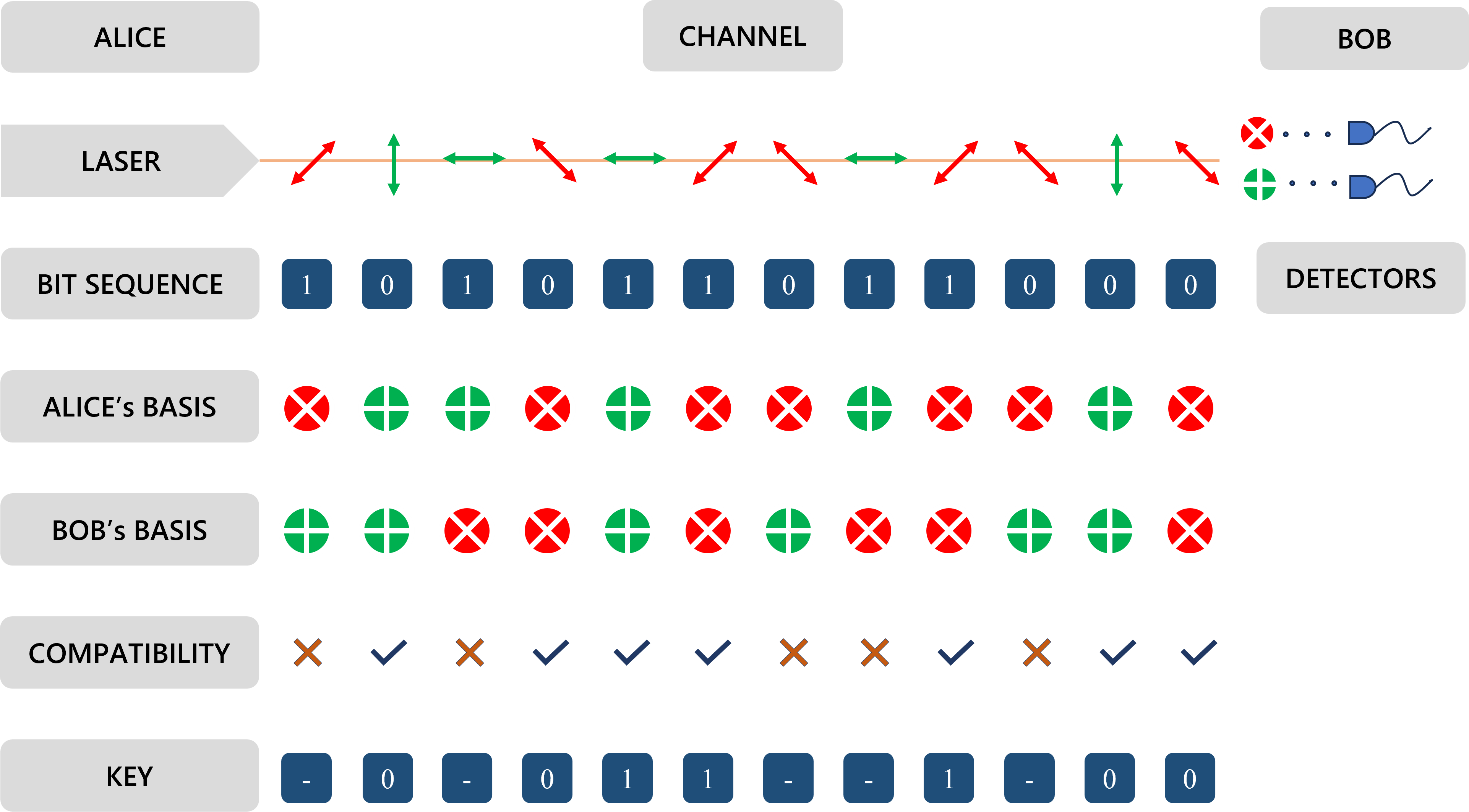} \caption{Schematic of the BB84 protocol depicting Alice's encoding choice and Bob's measurement choice. To obtain a correlated key, Alice and Bob only retain those results where their respective basis choices match.}
    \label{fig:bb84_scm}
\end{figure}

In the polarisation-based BB84 protocol, Alice encodes her bits using randomly chosen bases, either the rectilinear basis ({H, V} or 0°, 90°) or the diagonal basis ({D, A} or -45°, 45°), which are Mutually Unbiased Bases (MUBs). She prepares single photons in one of these polarisation states and transmits them to Bob via a quantum channel. Upon receiving the photons, Bob independently and randomly selects a basis to measure each one and records the outcomes. Alice and Bob then discuss their basis choices over a public channel, without revealing the measurement results. They retain results where their bases match, known as sifting, to form the key. When Bob chooses the incorrect basis, his measurement outcome is uncorrelated with Alice's bit; if he chooses the correct basis, he obtains the correct bit value as shown in Fig.\,\ref{fig:bb84_scm}. A portion of the key is then publicly announced to check for errors. If the error rate is below a threshold, the protocol is deemed successful, and the obtained key can then be used for further communication.

The secure key rate for the BB84 protocol, considering ideal sources and detectors, is given by
\begin{equation}
    R=[1-2H_{\mathrm{bin}}(e)], 
    \label{SKRI}
\end{equation}
where $R$ is the secret key rate, $e$ is the QBER, and $H_{\mathrm{bin}}$ is the binary entropy. 
\noindent
In practical implementations, ideal single-photon sources are not readily available. Therefore, attenuated laser pulses or weak coherent pulses (WCP) can be used that emit, on average, one photon per pulse. However, the Poissonian photon statistics of the laser open up avenues for photon number splitting attacks (PNS) \cite{lutkenhaus2002quantum}, where the presence of multiple photons in a pulse aids the eavesdropper in gaining some information on the shared secret key. 
A possible defence against PNS attacks is the use of decoy pulses \cite{ma2005practical}. By incorporating decoy pulses with varying intensities alongside the signal pulses, Alice and Bob can monitor discrepancies in the transmission statistics, hence protecting against PNS attacks. The secret key rate for the decoy state protocol is then given by
\begin{eqnarray}\label{eqn: decoy_DW rate}
R \geq \frac{1}{2} \left\{ - Q_{\mu} H_{\mathrm{bin}}(E_{\mu}) f(E_{\mu}) + Q_1 \left( 1 - H_{\mathrm{bin}}(e_1) \right)\right\}.
\end{eqnarray}
Here, $ \mu$ is the mean photon number of the signal pulses, $Q_\mu$ is the overall gain of signal states, $E_\mu$ is the overall QBER, $Q_1$ is the gain of single-photon states, and $e_1$ is the error rate of single-photon states.

A further avenue of attack on the system is through the non-idealities of the devices or through side channels \cite{nauerth2009information, arteaga2022practical}. Such flaws in the experimental implementations offer great potential for hacking the QKD system \cite{miller2025micius}. To accurately quantify the amount of information going to Eve through such means, it is essential to understand and account for the imperfections in the devices used. This requires thorough calibration of the equipment on both Alice’s and Bob’s sides, enabling the identification and quantification of any side-channel information that may be accessible to an adversary.  Each system component can be characterised to obtain an upper bound on the maximum distillable secure key. Our work quantifies certain information leakages arising from physical layer device imperfections and provides an estimated upper bound on the secret key rate. Although these leakages are not exhaustive and additional loopholes may exist, they offer a practical estimate of the distillable secure key achievable with real QKD devices. 

\subsubsection{\textbf{EPCD protocol}}

This protocol adopts a comprehensive strategy by utilising entrapped pulses in conjunction with coincidence monitoring. 

In the Entrapped Pulse Coincidence Detection (EPCD) protocol\cite{sharma2025enhancing}, Alice sends a signal pulse of mean photon number $\mu$ together with additional ``entrapped'' pulses of mean photon number $\nu$. At Bob’s end, we monitor the coincidences to examine whether the photon statistics have been altered or remain unchanged. Because Eve cannot discriminate between the signal and entrapped decoys, she cannot reliably mimic the coincidence statistics without introducing detectable discrepancies. If the detected coincidences match the expected statistics, contributions from both single- and two-photon pulses are included in the key rate calculation, thereby improving performance over the standard decoy-state method. The primary distinction lies in the coincidence monitoring, with
no signiﬁcant impact on the experimental feasibility of the protocol.

%In the EPCD protocol, Alice sends a signal pulse along with decoy pulses called entrapped pulses. These pulses help detect eavesdropping by monitoring coincidence events due to multiphoton pulses at Bob’s end. If the measured two-fold coincidences at Bob's side match the expected two-fold coincidences, it indicates the channel is secure. Any deviation shows the presence of an attack, such as a photon number splitting attack. This process is repeated for each transmission to ensure ongoing security during key generation. 

Due to the contribution of multiple photons, the equation for the secure key rate becomes:
\begin{eqnarray}
R \geq && \frac{1}{2} \Big\{ - Q_{\mu} H_{\mathrm{bin}}(E_{\mu}) f(E_{\mu}) + Q_1 \left( 1 - \Phi(2 e_1 - 1) \right)\nonumber \\  
 &&+ Q_2 \left( 1 - \Phi((2 e_2 - 1)^2) \right) \Big\},
\label{eqn:DW_rate2}
\end{eqnarray}
where $\Phi(x):= H_{\mathrm{bin}}(1/2 + x/2)$, $Q_2$ is the gain of two-photon states, and $e_2$ is their corresponding error rates. The key rate is then computed via a Semi-Definite Programming (SDP) based optimisation method, whose details can be found in Ref.\,\cite{sharma2025enhancing}.

\section{\label{sec:pre}Experimental Prerequisite}

Before starting the experiment, we follow certain steps to ensure the effective implementation of the QKD protocol. These preparatory steps are crucial for minimising errors and enabling secure key generation. These steps include source, channel, and detector characterisation along with time synchronisation between transmitter and receiver.

\subsection{\textbf{Source characterisation}}

For implementing a polarisation-based BB84 protocol, the source can be designed in two ways. First, using four laser diodes for encoding the polarisation qubits, where each laser is randomly triggered to prepare the corresponding state. Second, using one or two lasers and employing active electro-optic modulators to encode the state \cite{Avesani:21}. The advantage of the latter method is the reduction in complexity of the system, with a trade-off in transmission rate limited by the bandwidth of the modulators. The former method typically enables a higher key generation rate, only limited by the laser drivers. However, it leaves the system open to side-channel attacks, which arise from the laser diodes being non-identical \cite{ko2017critical}. 

\begin{figure}[h]
    \centering
    \includegraphics[width=11cm]{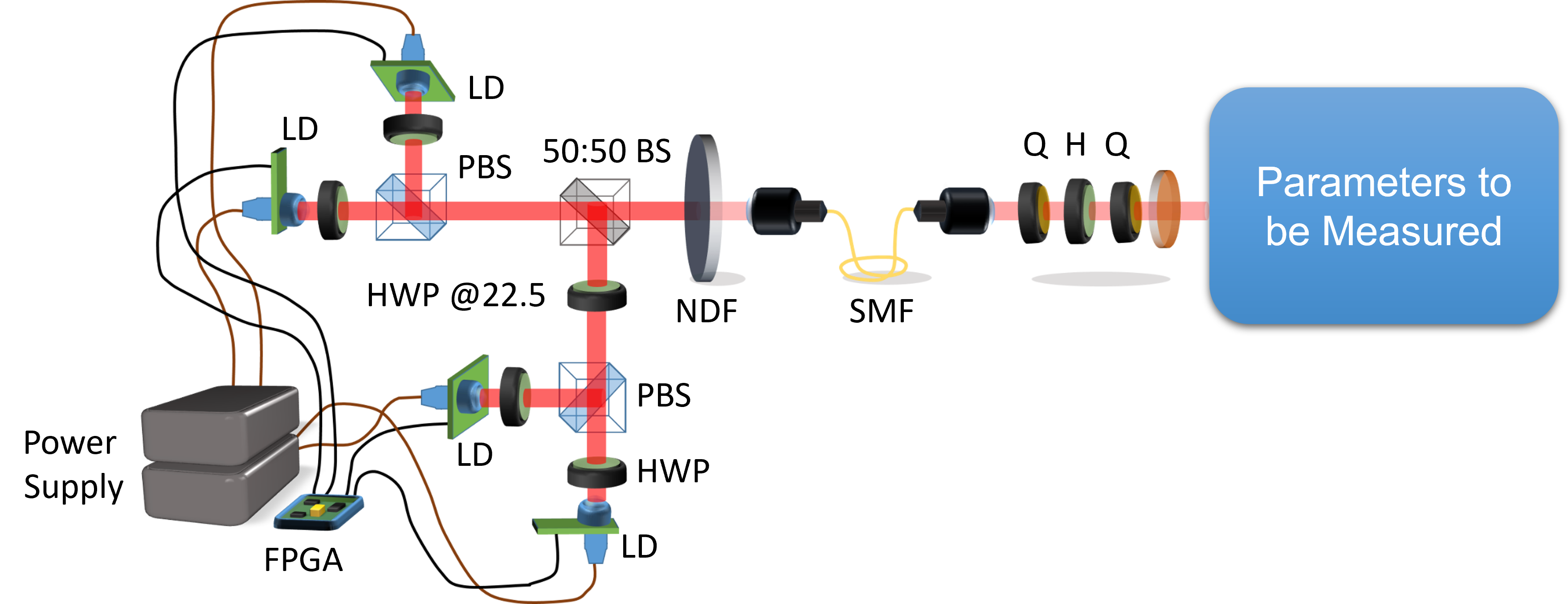}
    \caption{Schematic of experimental setup for source characterisation. It includes both the optics and electronic components. FPGA: Field Programmable Gated Array, LD: Laser Diodes, HWP: Half Wave Plate, BS: Beam Splitter, PBS: Polarising Beam Splitter, NDF: Neutral Density Filter, SMF: Single-Mode Fibre, Q: Quarter wave plate, H: Half wave plate}
    \label{fig:source_char}
\end{figure}

\begin{figure}
     \centering
     \begin{subfigure}[b]{0.4\textwidth}
        \centering
         \includegraphics[width=5.1cm]{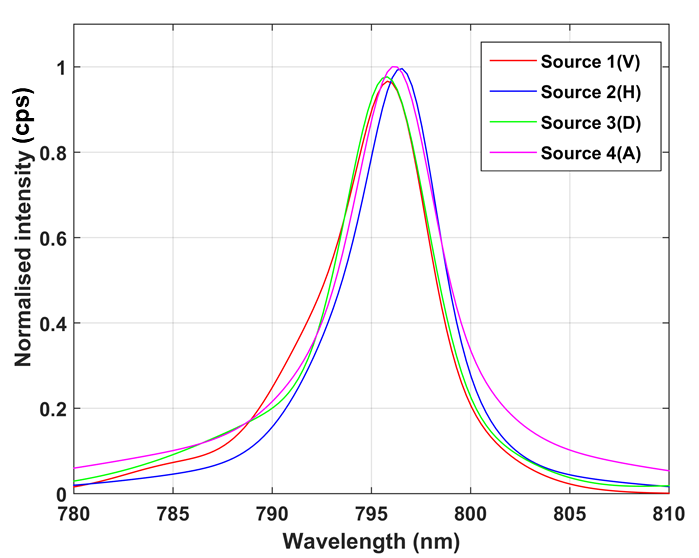}
         \caption{\footnotesize Wavelength of the four laser diodes, recorded using a spectrometer.}
         \label{fig:source_wave}
     \end{subfigure}
     \hspace{1cm}
     \begin{subfigure}[b]{0.4\textwidth}
         \centering
         \includegraphics[width=5cm]{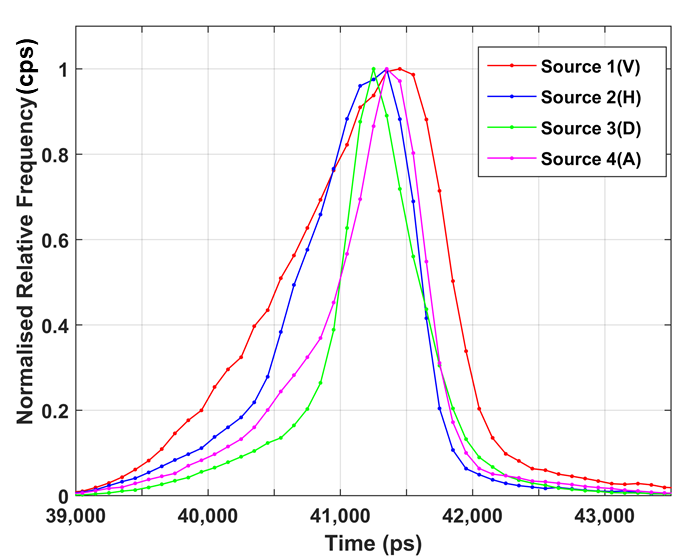}
         \caption{\footnotesize The photon arrival time from four laser diodes, recorded using SPAD.}
         \label{fig:source_pulse}
     \end{subfigure}
     \vfill
     \begin{subfigure}[b]{0.45\textwidth}
         \centering
         \includegraphics[width=5.4cm]{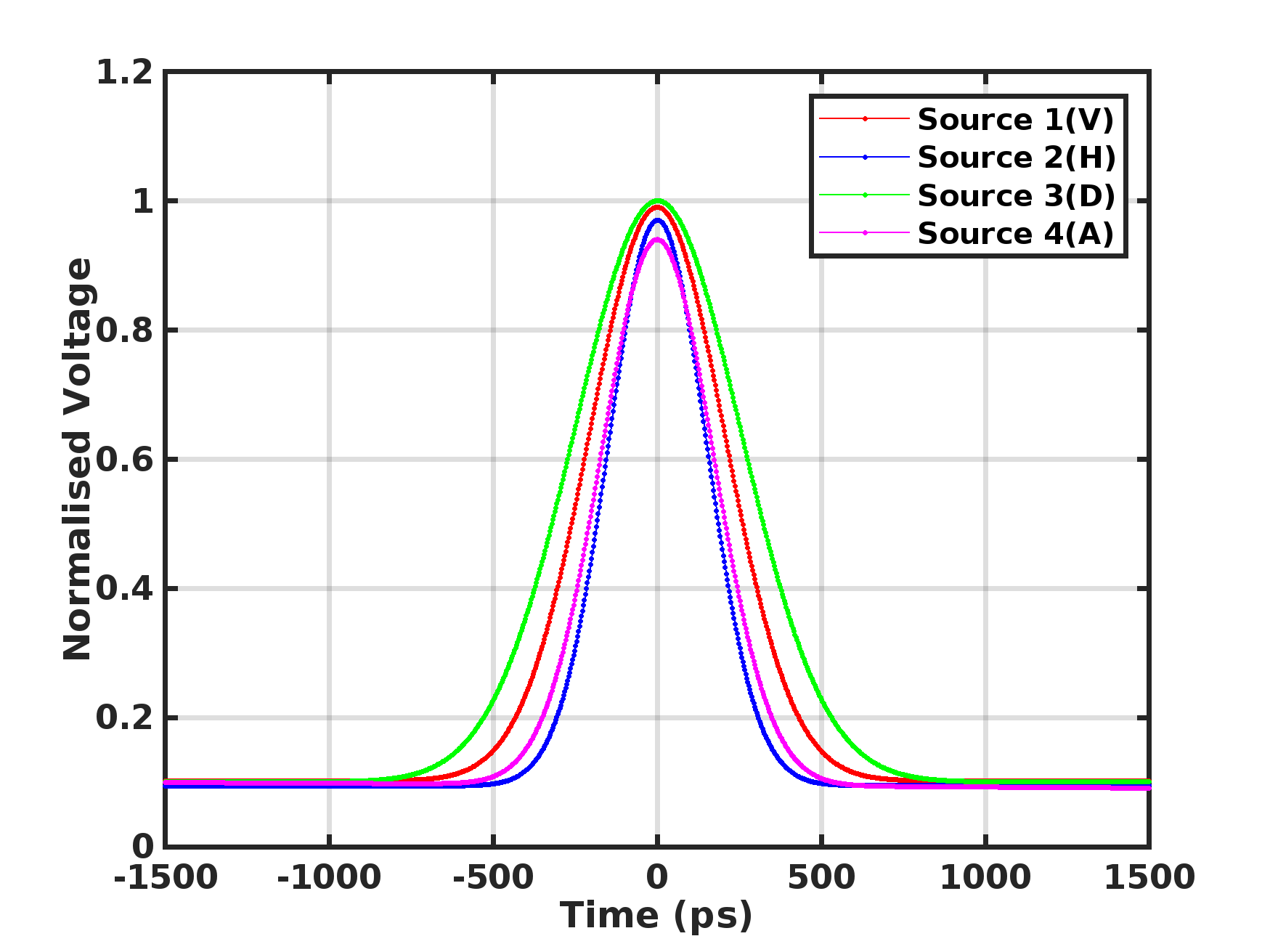}
         \caption{\footnotesize Optical pulse width of four laser diodes, recorded using a photodiode and an oscilloscope.}
         \label{fig:source_time}
     \end{subfigure}
        \caption{ Source parameters used for characterising the states prepared by Alice.}
        \label{fig:sources_all}
\end{figure}

In our Free-Space QKD system, we have opted for the former method of using four laser diodes to encode the polarisation qubits. To protect against side-channel attacks on the source end, we have conducted a comprehensive characterisation of these lasers, considering parameters such as spectral width, pulse width, spatial mode, peak wavelength, polarisation, and arrival times at the receiver. The condition of security under the absence of side-channels is given by 
\begin{equation}
    \rho_{S_i}(\lambda_i, w_i, \overrightarrow{S_i}) = \rho_{S_j}(\lambda_j, w_j, \overrightarrow{S_j}).
\end{equation}
Here, $\rho_{S_i(S_j)}$ denotes the quantum state emitted by source $S_i (S_j)$; the above equation demands that the states emitted by the different sources are identical in all degrees of freedom.
Rigorous characterisation allows us to assess potential information leakage through side channel attacks by analysing the mutual information between the information source and a potential eavesdropper. The experimental setup for performing such a characterisation is as shown in Fig.\,\ref{fig:source_char}. Side channel attacks occur when Eve takes advantage of implementation imperfections to gain knowledge about the source or detector, which are not observed through QBER. The amount of information leakage due to the laser diodes being non-identical is given by  
\begin{equation}
    \label{eq:6}
    I(A:E) = 1 + \sum_{a \in \mathscr{E}}\sum_{b \in \mathscr{B}} \frac{p(a|b)}{2} \log_2\left(\frac{p(a|b)}{2 p(a)}\right).
\end{equation}
In Eq.\,\ref{eq:6}, $b$ is the outcome of Eve’s measurement in her
device. $p(a)$ denotes the probability distribution over a given source parameter $a$, which can represent source parameters like wavelength, pulse duration and photon launch time. Any of these parameters may exhibit imperfections. When Eve observes an outcome $b$ (such as a polarisation state in the BB84 protocol), the conditional probability $p(b|a)$ expresses the chance that this outcome corresponds to a particular value of the parameter $a$.
The experimental results of the source characterisation can be seen in Figs\,\ref{fig:source_wave}-\ref{fig:source_time}. The information leakage due to wavelength, pulse width, and arrival time mismatch is estimated to be 0.001 bits/pulse. 
Drawing from our experimental findings regarding the cross-correlation between parameter values across different laser diodes, we propose methods to mitigate information leakage to a potential eavesdropper \cite{source}. 

Another aspect to be considered on the source side is photon statistics and the mean photon number. While single-photon-source (SPS) based QKD experiments have been performed \cite{basso2021quantum, zahidy2024quantum}, they particularly suffer from low key rate and complicated experimental setups. The alternative is to use attenuated laser pulses called weak coherent pulses (WCP). WCP used in QKD follows a Poisson distribution. Inaccurate estimation of the mean photon number can result in unaccounted information leakage due to multi-photon pulses, potentially enabling undetected adversarial attacks. Therefore, precise characterisation of the mean photon number is essential in practical QKD systems based on WCP. Since the single-photon avalanche diodes (SPADs) used are not photon-number resolving, using multiple on-off detectors provides improved resolution. 

We initially estimate the average photon number using a single detector. For a more precise value, we then adopt the four-detector characterisation approach from a previous study \cite{kumazawa2019rigorous}. This advanced setup accurately determines the Poisson statistics of the WCPs by providing precise probabilities for lower photon numbers.
This experiment allows us to quantify the deviation and analyse the resulting information leakage caused by the inaccurate calculation of mean photon number \cite{sharma2024mitigating}.
\noindent
For a source, which consists of pulsed lasers with a repetition rate of $\nu_{\mathrm{rep}}$, emitting at a wavelength $\lambda$ and emitting with an average power of $P_{\mathrm{avg}}$, the mean photon number per pulse ($\mu$) is given by
\begin{equation}
    \mu = \frac{P_{\mathrm{avg}}\lambda 10^{-OD}}{\nu_{\mathrm{rep}}h c},
\end{equation}
where OD is the Optical Density of the employed neutral density filters. The desired mean photon number can then be achieved by employing various neutral density filters and attenuators.

\subsection{\textbf{Channel Characterisation}}

In free-space QKD, accurate channel characterisation is essential for secure and reliable key exchange. At distances like 200 meters, the channel is primarily influenced by atmospheric absorption, and scattering depends mainly on weather and ambient conditions. Atmospheric turbulence can introduce beam wander, scintillation, and wavefront distortions, affecting signal stability and increasing the QBER \cite{heim2010atmospheric, vasylyev2018characterization}.
The distance and extinction coefficient of atmospheric aerosols equally affect the transmissivity of the channel. Higher aerosol extinction generally leads to a lower key rate \cite{mishra2022bbm92, jabir2017robust}. The atmospheric transmission is given by
\begin{equation}
    T = S_{C} e^{-\gamma L},
\end{equation}
where $S_{C}$ is a scaling parameter, $\gamma$ is the extinction coefficient and $L$ is the channel length.
In polarisation-based BB84, preserving polarisation integrity is crucial, as depolarisation effects or optical misalignment can lead to a high QBER, further affecting the secure key rate.
In our implementation, we have considered transmission loss, path delay and atmospheric conditions as they are crucial for accurately representing the quantum channel and ensuring the optimisation of free-space quantum communication systems' performance.

\subsection{\textbf{Detector Characterisation}}

For discrete-variable QKD, single-photon detectors are crucial for detection at the receiver. Commercially available superconducting nanowire single-photon detectors (SNSPDs) offer high efficiency and low noise but require cryogenic temperatures, limiting their use in field implementation. Therefore, we focus on room-temperature single-photon avalanche photodiodes (SPADs). Silicon-based SPADs in the visible spectrum offer good efficiency (62\%) and a saturation rate over 5 MHz. For this reason, we use the Excilitas SPCM-AQRH FC module with four detectors, suited for the polarisation-based BB84 protocol. It has a peak quantum efficiency at 650 nm and typically consumes 3.3 W, with power mainly used for the thermoelectric cooler that reduces dark counts.

\begin{figure}[h]
    \centering
     \includegraphics[width=8cm]{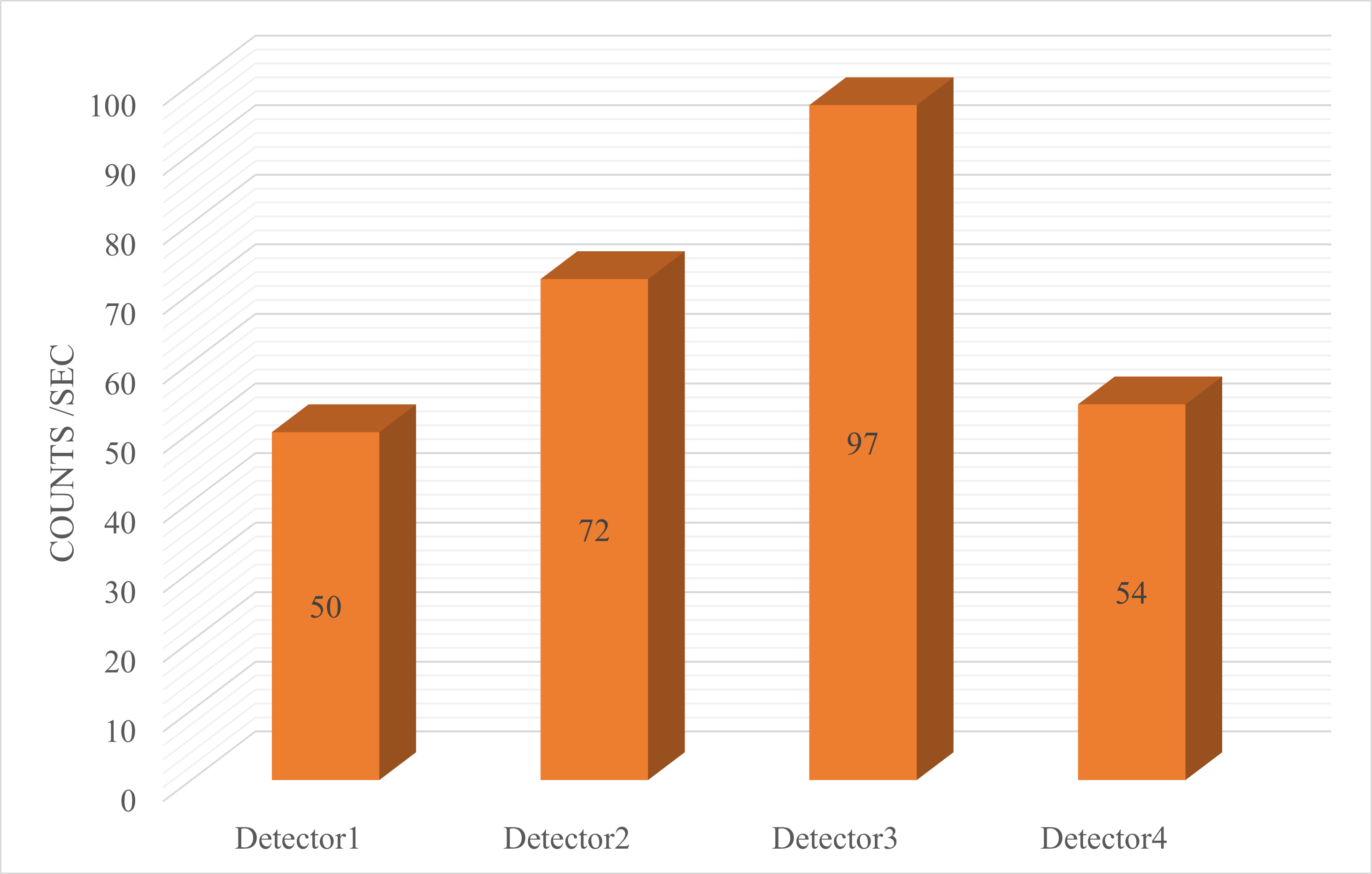}
    \caption{Dark counts of the single-photon avalanche photo-diodes (SPADs).}
    \label{fig:darkcount}
\end{figure}

Ideally, as per the specification sheet, all detectors are assumed to have identical detection efficiencies. However, in practical implementations, detector efficiencies may exhibit mismatches due to variations in manufacturing or experimental setup. Fig.\,\ref{fig:darkcount} shows the measured dark count rates for each of the four SPADs used in the experiment. Additionally, such mismatches can arise from the differential coupling of various spatial modes of incoming signals into the detectors \cite{zhang2021security}.
In our work, we also consider the effects of coupling mismatches between detectors at the receiver's end. Specifically, we analyse the potential information leakage to an eavesdropper (Eve) resulting from such mismatches, quantified in terms of the mutual information between the eavesdropper and the receiver \cite{sharma2023vulnerability}.

\subsection{\textbf{ Time Synchronisation between Transmitter and Receiver}}

Synchronisation between the transmitter and receiver is crucial for effective quantum communication, ensuring accurate transmission and detection of quantum states. Both the transmitter (Alice) and receiver (Bob) must align their clocks precisely in order to perform classical post-processing correctly. Timing errors can result in missed or incorrect measurements, increasing the QBER. Achieving synchronisation typically involves methods like Network Time Protocol (NTP). Pulse-based synchronisation and feedback loops are also employed to maintain timing. However, challenges like noise in the quantum channel, the need for high precision over long distances, and ensuring timing accuracy must be addressed. 

Given that the transmitter (Alice) and receiver (Bob) are situated at a distance from each other and operate with independent reference clocks, achieving precise timing synchronisation is of utmost importance before labelling the photon pulse sequence for key sifting and correlation. To address this requirement, a synchronised time base is imperative at both ends of the communication \cite{clocksynch}.

The primary goal of the synchronisation procedures is to establish a clear correspondence between photon detection events at the receivers and photons emitted from the source. The aim is to maintain an automated framework that consistently generates secure keys without errors.

Since Alice and Bob's terminals remain stationary in our case, the time-of-flight for the photons remains constant. However, two major challenges need to be addressed in QKD timing synchronisation:
\begin{enumerate}
    \item 
 Clock Synchronisation (Temporal alignment of Tx and Rx clock)
\item
Data frame alignment (Precise time tagging)  
\end{enumerate}

For clock synchronisation, we have used reference clocks within the Field-Programmable Gate Arrays (FPGAs) located at both ends of the communication system. To achieve synchronisation, we have established time-based synchronisation by transmitting the reference clock of 100 MHz from Alice's FPGA to Bob's FPGA, thereby ensuring precise synchronisation of time bases for precise time-tagging during photon detection events.
In the context of frame synchronisation, we ensured that the initiation of photon generation at the transmitting end (Alice) and the counting process at the receiving end (Bob) are synchronised. This is accomplished by sending one Pulse Per Second (PPS) pulse from Alice to Bob and is referred to as absolute time tagging, as they serve to synchronise the primary starting point of the time counters involved in these operations.
When the transmitter and receiver are in motion or separated by large distances, such as in satellite-based QKD or long-distance QKD over optical fibers, clock synchronisation is typically achieved using Global Positioning System (GPS) signals or through classical communication channels over the same fiber link \cite{erven2008entangled,berra2023synchronization}. In these scenarios, increased clock jitter becomes a significant concern and must be carefully addressed to maintain precise timing and synchronisation.

%%%%%%%%%%%%%%%%%%%%%%%%%%%%%%%%%%%%%%%%%%%%%%%%%%%%%%%%%
%                ~Experimental Method~                  %
%%%%%%%%%%%%%%%%%%%%%%%%%%%%%%%%%%%%%%%%%%%%%%%%%%%%%%%%%

\section{QKD Subsystems}\label{Sec: practical}
A typical P$\&$M QKD system comprises a transmitter that randomly prepares quantum states and sends them through a free-space channel to a receiver, which performs the corresponding measurements. The following subsections detail the experimental implementation of the transmitter, channel, and receiver, together with the data acquisition and automation of protocol execution, essential for reliable and fast operation.

\subsection{Transmitter}

In Alice's terminal, as illustrated in the Fig.\,\ref{fig:Alice}, the transmitter incorporates an in-house developed pulse laser diode driver module. The random triggering of these laser modules is achieved through random transistor-transistor logic (TTL) voltage pulses generated using an FPGA. The input beam traverses through a combination of Half Wave Plates (HWPs) and Polarising Beam Splitters (PBSs), which perform the dual duty of both controlling the mean photon number and the polarisation. Before launching the quantum signal from Alice to Bob, the beam is directed to the launching optics, where it is magnified, collimated, and then launched to Bob. The experiments were performed in an approximately 200-meter free space channel.
 
\subsubsection{\textbf{Laser diode driver module}}

The implementation of P$\&$M QKD protocols like BB84, B92, etc requires the operation of laser diodes with high repetition rates and short pulse widths to prepare quantum states at a high rate. For performing QKD experiments, laser diode parameters like optical power and pulse width should be adjustable and easy to control. We have constructed a laser diode driver circuit that drives laser diodes and is used in the QKD transmitter as shown in Fig.\ref{fig:Alice}. This driver circuit can drive the laser diodes up to 5 MHz with a pulse width of around 1 ns \cite{uhring2004low}. The repetition rate and pulse width are constrained by the PCB design and the switching characteristics of the laser diodes. This circuit needs two inputs: one is a voltage supply to provide sufficient current to the diode, and the other is a trigger input, which is a TTL pulse to control the switching operation of the diode. Based on the TTL voltage, the diode can be switched on and off as required.

\subsubsection{\textbf{RNG for enabling laser diodes}}

In QKD, using a high-quality Random Number Generator (RNG) can play a vital role in ensuring efficient and secure key generation. Ideally, QRNG \cite{mongia2024investigating} is used for QKD; however, for practical implementations, Pseudo-Random Number Generators (PRNGs) can meet the randomness requirements. While PRNGs are deterministic, their security depends on the quality of the seed and the entropy used to initialise them. When combined with other security measures like post-processing techniques, a good-quality PRNG can contribute to better security, providing an efficient solution while maintaining the integrity of the cryptographic system \cite{akhshani2014pseudo}. 

\begin{figure}[h]
    \centering
     \includegraphics[width=8cm]{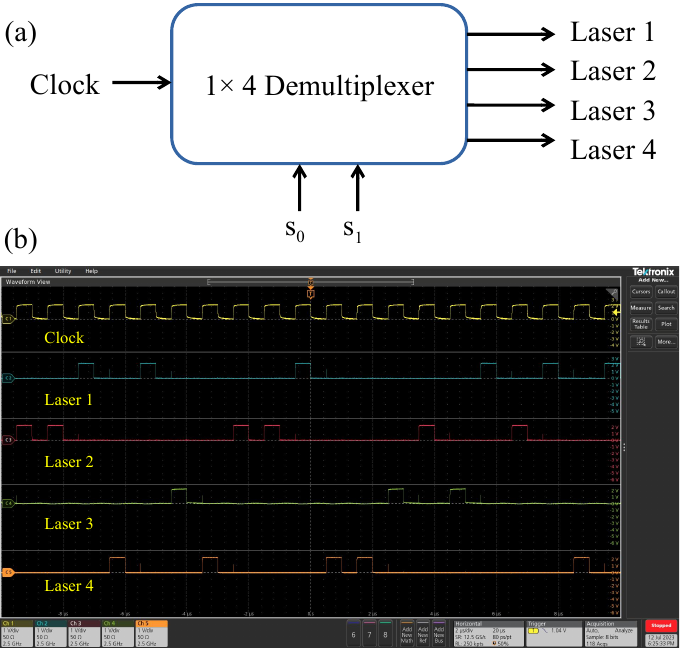}
    \caption{a) Schematic of generating four TTL pulses using a 1$\times$4 demultiplexer using the random bits generated from LFSR as a select line $s_0$ and $s_1$. The select line enables one laser at a time and performs the demultiplexing. b) Four colours representing four random TTL pulses generated from the FPGA to drive four corresponding laser diodes, each diode corresponds to polarisation states ($H$, $V$, $D$, $A$).}
    \label{fig:RNG}
\end{figure}

In our implementation, a linear feedback shift register (LFSR) based PRNG with XOR technique was developed on an FPGA to generate random TTL pulses for driving the laser diodes. We generated two different random sequences using different length seeds using LFSR and combined them using a bitwise XOR operation. While each individual sequence failed the NIST statistical test suite, the XOR-combined output passed all tests, indicating enhanced randomness, confirming its suitability for cryptographic applications \cite{chandravanshi2023lfsr}. In the BB84 implementation, to ensure that the four laser diodes used for state preparation are triggered randomly, we have implemented a de-multiplexer in the FPGA, as shown in Fig.\,\ref{fig:RNG}. The input of this de-multiplexer is the clock pulse, and the RNG bits are used as a select line. Based on the select line, at a time only one laser diode will get a trigger pulse, and the others will be disabled as shown in Fig.\,\ref{fig:RNG}. 

\subsubsection{\textbf{Polarisation encoding optics}}

\begin{figure}[h]
    \centering
     \includegraphics[width=12cm]{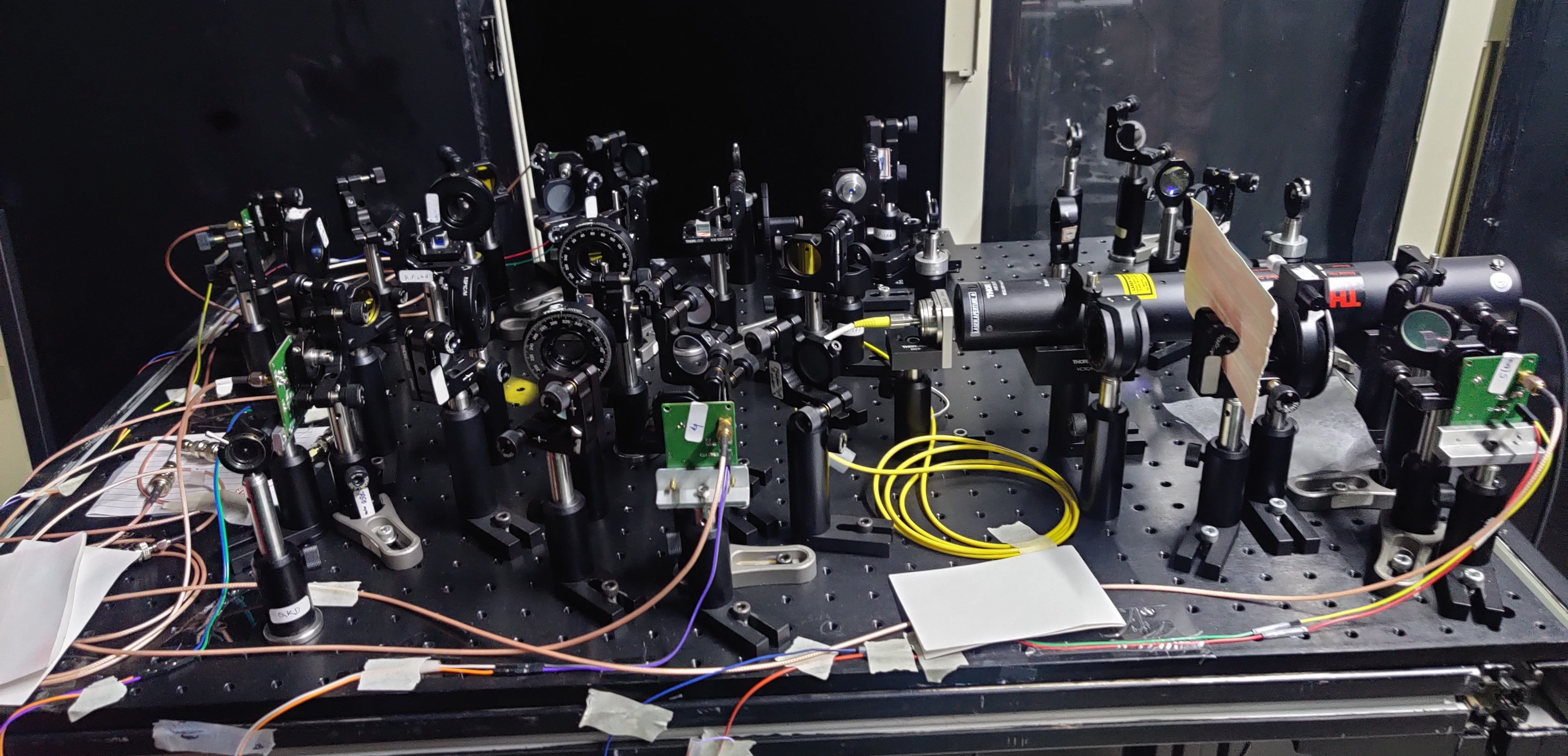}
    \caption{Polarisation encoding optics with launching optics at Alice, it consists of all four lasers with one He-Ne beacon laser. All beams are combined and launched to a free space channel. The size of the optical breadboard is 90 cm$\times$60 cm.}
\label{fig:Alice}
\end{figure}

Alice encodes her quantum states in the polarisation degree of freedom using weak coherent pulses. These pulses are generated by attenuating laser outputs with neutral density filters. Our optical setup, Fig.\,\ref{fig:Alice}, consists of four Thorlabs $L808P010$ laser diodes operating at 808 nm. The diodes are driven by a custom-designed laser driver circuit controlled via an FPGA, producing optical pulses at a 5 MHz repetition rate with a pulse width of 1 ns.
Four laser outputs are engineered to produce the polarisation states horizontal (H), vertical (V), diagonal (D), and anti-diagonal (A) using a combination of polarising beam splitters (PBS) and half-wave plates (HWP). This combination performs the dual duty of both controlling the mean photon number and the polarisation. The PRNG on the FPGA ensures unbiased triggering of the lasers, enabling random generation of the four polarisation states.

After combining individual laser pulses using a beam splitter (BS), the signal passes through one fixed and one variable neutral density filter (NDF) for further attenuation. The variable NDF is adjusted to tune the intensity of the pulses, which is calibrated in terms of the mean photon number.

\subsubsection{\textbf{Launching Optics}}

Before launching the signal through the free-space channel, the laser pulses are combined with a beacon laser to facilitate alignment of the transmitter. In our setup, a 633 nm He-Ne laser serves as the beacon, making alignment straightforward. In order to collect the maximum amount of photons at the receiver, beam divergence has to be minimised. Since beam divergence is inversely related to the beam diameter, the signal beams are expanded to reduce beam divergence. The beam expansion is realised using two plano-convex lenses in the 4-f configuration. The lenses employed in our setup are a lens with a 5 cm focal length of 1-inch diameter, followed by a lens with a 50 cm focal length of 2-inch diameter.

\subsection{Channel between Alice and Bob}

Our communication infrastructure is established across two proximate buildings within the Thaltej campus of the Physical Research Laboratory (PRL). The transmitting and receiving stations are situated on the same rooftop but in separate rooms.

\begin{figure}[h]
    \centering
     \includegraphics[width=8cm]{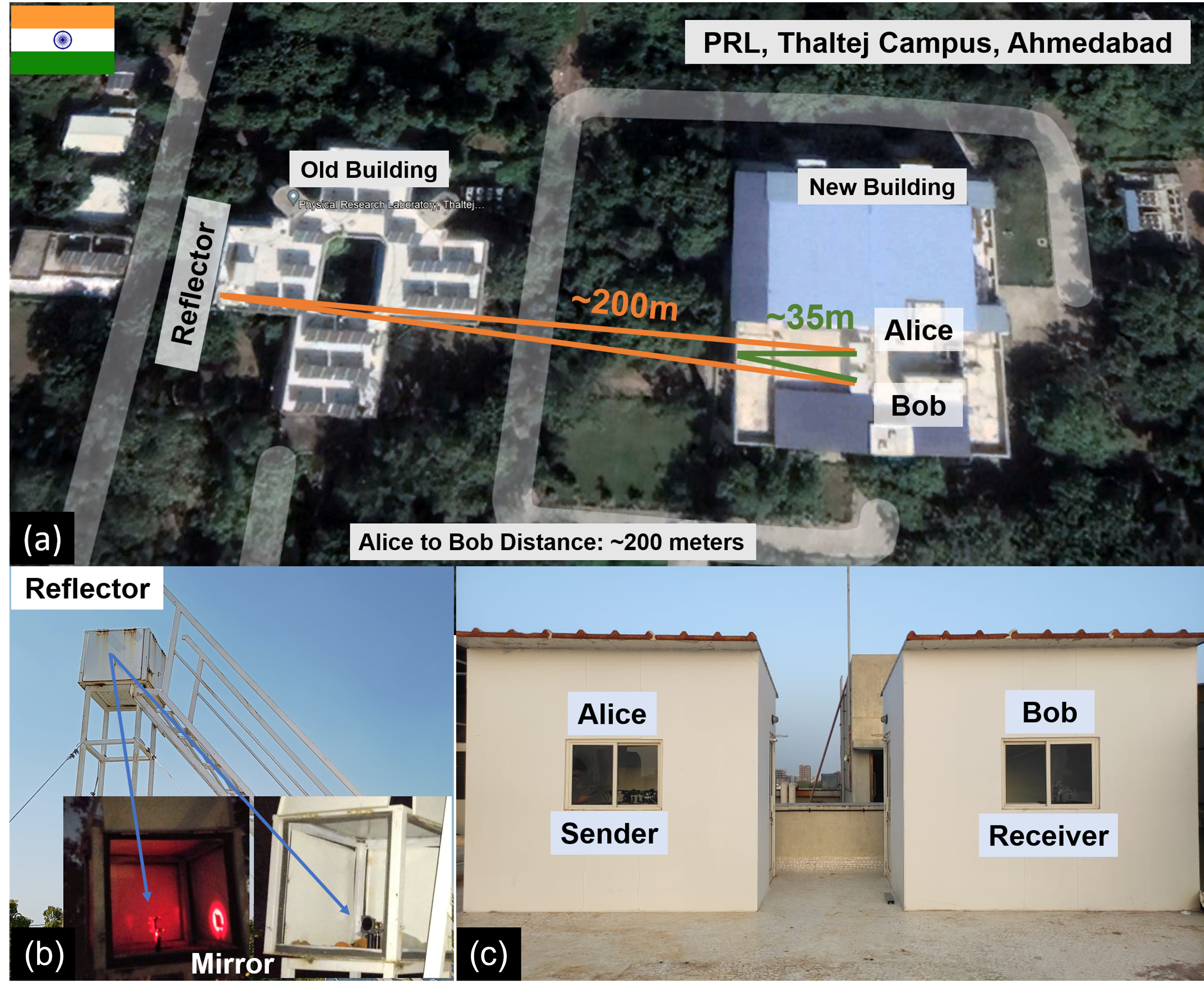}
    \caption{Arrangement of various components in the channel. It includes the location of Alice and Bob, and their setups. (a) Aerial view of free space quantum channel (Image courtesy: Google Earth), (b) Reflector consists of one 2-inch aperture mirror, (c) front view of Alice and Bob's room.}
    \label{fig:terrace}
\end{figure}

In our experiment, the quantum channel is a free space optical channel, and the classical channel is an Internet/Ethernet interface. To facilitate quantum communication over 200 meters, we use a reflector on the adjacent building, which utilises a 2-inch diameter mirror mounted on a dedicated structure. The building structure and equipment placement are depicted in Fig.\,\ref{fig:terrace}. Alice is situated in one cabin and is responsible for generating and transmitting polarisation states, while Bob occupies an adjacent room and detects the incoming photons.

\subsection{Receiver}

Bob's terminal is equipped with collecting optics and polarisation decoding optics. The incoming photons are subsequently decoded and detected by Single Photon Avalanche Detectors (SPAD), and the detection events are recorded by the Rx-side FPGA.
\noindent
The alignment process commences with coarse alignment using a visible beacon laser operating at a wavelength of 633 nm at Alice's side.

\subsubsection{\textbf{Collection Optics}}

\begin{figure}[h]
    \centering
     \includegraphics[width=8cm, height=5cm]{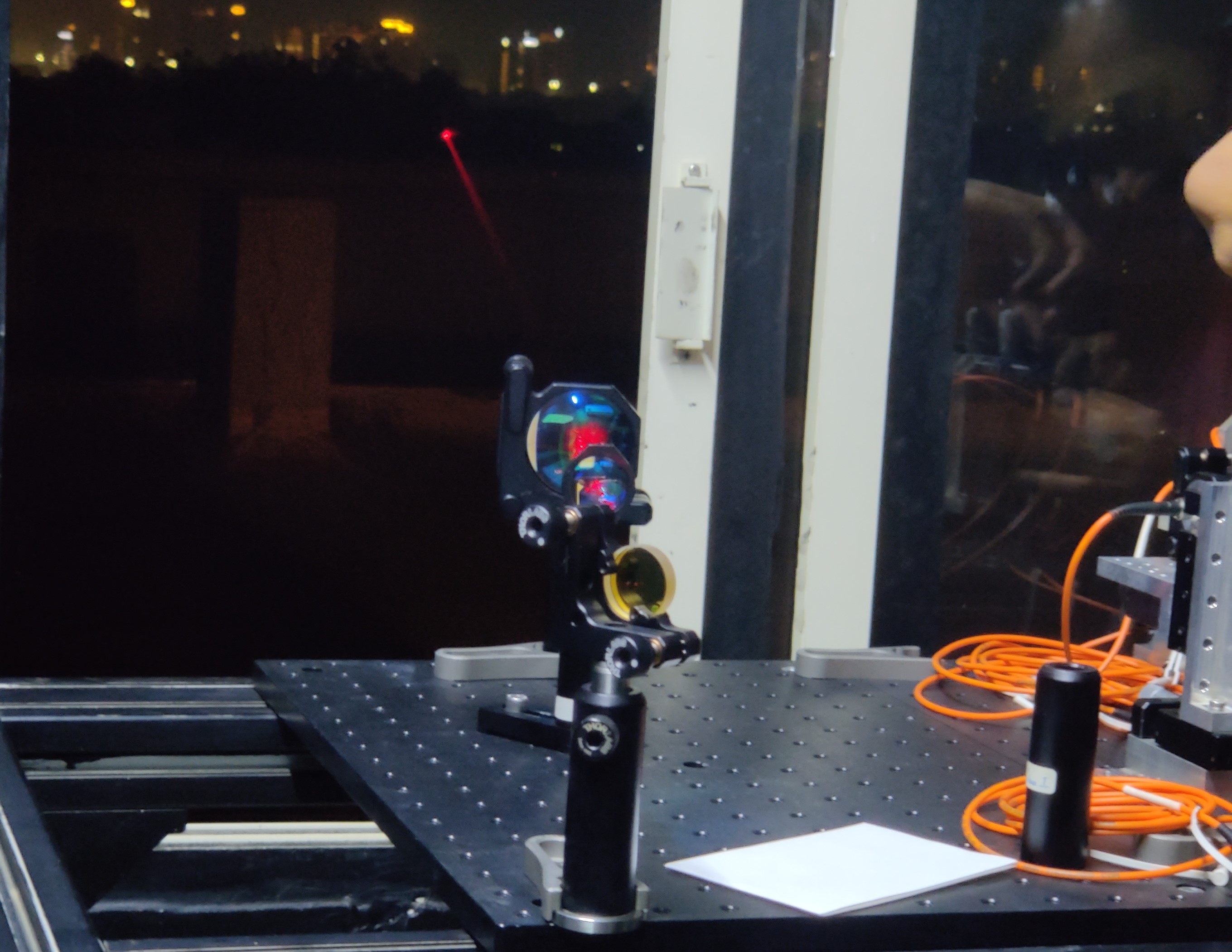}
    \caption{Collection optics at Bob consisting of a combination of two lenses.}
    \label{fig:collection}
\end{figure}

At the receiving end, the incoming signal is initially gathered using a similar lens setup as in the transmitter, but in an opposite configuration to reduce the beam size. It consists of a lens with 50 cm focal length of 2-inch diameter and a lens with 5 cm focal length of 1-inch diameter as shown in Fig. \ref{fig:collection}.

\subsubsection{\textbf{Polarisation decoding optics}}

\begin{figure}[h]
    \centering
\includegraphics[width=8cm]{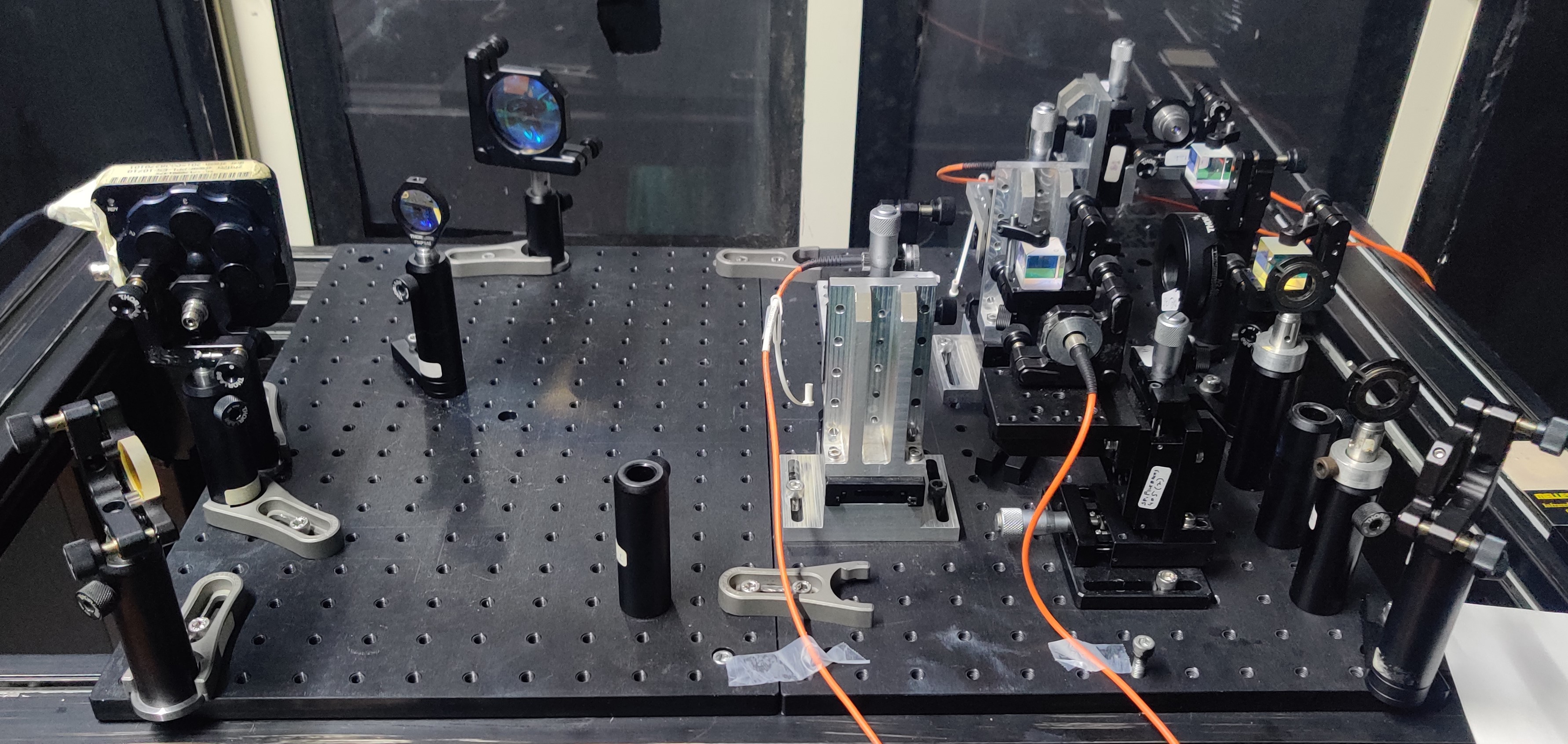}
    \caption{Polarisation decoding optics at Bob includes BS, PBS, HWP and other optics for aligning the beam to the detectors. All four polarised lights are coupled to four single-photon detectors using multi-mode fibres.}
    \label{fig:Bob}
\end{figure}

Once the collimated signal is obtained from the collection optics, it is directed towards Bob's polarisation decoding optics as mentioned in Fig.\ref{fig:Bob}, which include a beam splitter (BS) which performs the basis selection. 

The output from the beam splitter is sent through a polarising beam splitter (PBS) for measurement in the H/V basis, and a half-wave plate and PBS for measurement in the D/A basis. The decoded photons are then guided to single photon detectors (Excelitas SPCM-AQRH-14-FC)  through multi-mode fibres (MMF) for increased collection efficiency across all modes. The detector's outputs are connected to an FPGA-DAQ system, which records detection events and their timing information.

\subsection{Interfacing and automation }

Interfacing and automation of the transmitter and receiver DAQs are essential in QKD for efficient protocol execution. Automation manages real-time control, optimises performance, and handles tasks like state preparation, synchronisation, temporal filtering, parameter estimation, error correction, and privacy amplification, ensuring a secure final key generation.
This comprehensive software solution encompasses both the back-end and front-end components, effectively managing synchronisation, device control, and real-time protocol execution. Our implementation leverages the LabVIEW-FPGA platform along with Python to facilitate this, with the creation of a graphical user interface (GUI) panel.
\begin{figure}[h]
    \centering
     \includegraphics[width=8cm]{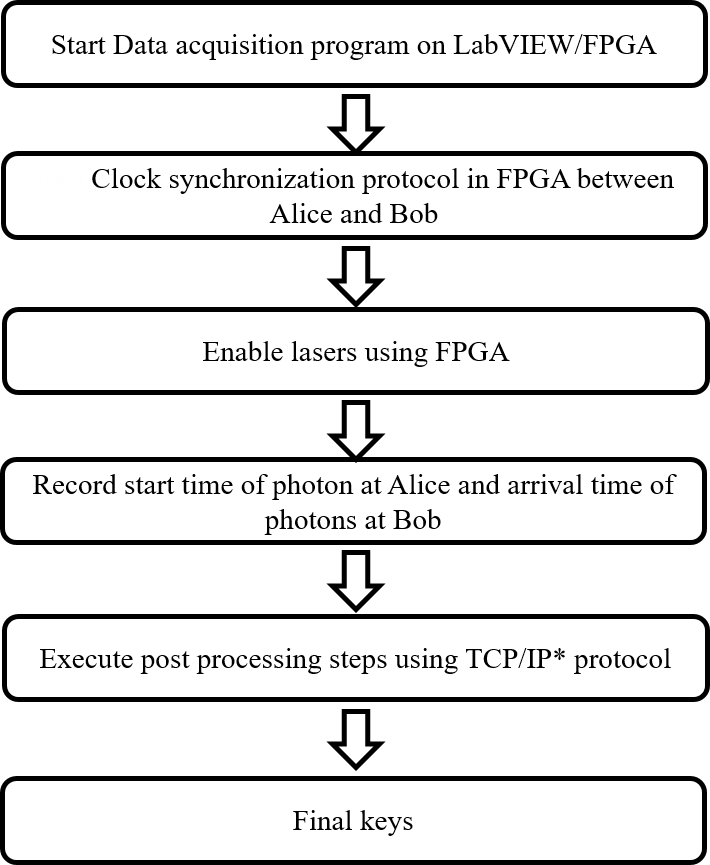}
    \caption{ Flow chart of real-time protocol execution. * Instead of TCP/IP, one can use other classical communication protocols for post-processing.}
    \label{fig:flowchart}
\end{figure}
The GUI panel serves the dual purpose of configuring the system settings within the QKD framework and providing real-time data logging functionality for all performance parameters. One can also use end-to-end VHDL/Verilog-based design in an FPGA for real-time post-processing protocols \cite{shingala2022real}.
The protocol execution steps are implemented on an FPGA board. We used a PXIe-7972 FPGA in combination with the NI-9478 digital I/O card for both the transmitter and receiver modules. All input and output channels are fully programmable. At the transmitter, the output ports are used to control and trigger the lasers, while at the receiver, four input ports are dedicated to acquiring TTL signals from single-photon detectors. For classical post-processing, a TCP/IP communication protocol is established between the transmitter and receiver. This hardware platform was chosen for its high configurability and seamless integration with LabVIEW, enabling efficient and flexible system control.
The flow of protocol execution is given in Fig.\,\ref{fig:flowchart}.

\section{Field implementation}\label{Sec:field}

We have implemented BB84 , BB84 with decoy\,\cite{ma2005practical},  coincidence detection (CD) protocol\,\cite{biswas2022quantum} and EPCD protocol\,\cite{sharma2025enhancing} in our field implementation. The technical details provided in this article are mostly focused on BB84, as the other above-mentioned QKD protocols are fundamentally derived from it. The details of a comparative study of all these protocols are given in \cite{sharma2025enhancing}.

\subsection{Real-time protocol execution}

The inter-building quantum channel of a 200-meter free space channel is established, and once the transmitter and receiver terminals are properly aligned, the experiment is conducted. After alignment, the FPGA initiates the predefined sequence of operations, culminating in the execution of the QKD protocol. The sequence of operations performed by the DAQ system is shown in Fig.\ref{fig:flowchart}.

\subsection{Post processing}
 
\subsubsection{\textbf{Sifting Algorithm}}
\label{subsubsec: sifting}

The sifted key rate can be calculated using the formula:

\begin{equation}\label{eq:siftkey}
n = 0.5 \ \nu_{rep} \ {\mu} \ \eta_{ch} \ \eta_{c} \ \eta_{d},
\end{equation}
where $n$ is sift key bits, $\nu_{rep}$ is repetition rate of laser, $\mu$ is mean photon number, $\eta_{ch}$ is channel transmission, $\eta_c$ is coupling efficiency, and $\eta_d$ is detector efficiency.

Using Eq.\,\ref{eq:siftkey}, the expected sifted key length can be estimated based on the known device and channel parameters. However, to extract the actual sifted key from the experimental data, the recorded time stamps at Alice and Bob are used. In this process, only the detection events where both parties have chosen matching bases are retained, while those with mismatched bases are discarded. This process relies on precise timing information, transmission time of photons from Alice, the corresponding detection time at Bob, and the propagation time through the quantum channel. We recorded the transmission timestamps at Alice and the detection timestamps at Bob, which are crucial inputs for the sifting procedure.
To facilitate efficient sifting, we have developed and implemented an algorithm for our QKD system based on the following equation. 
\begin{eqnarray}
t_{b_i} = t_{a_0}+ t_d + n_iT \pm T_f,
\label{eq:sifting}
\end{eqnarray}
where $t_{b_i}$ denote the time stamps of detection at Bob, $t_{a_0}$ denotes the start time at Alice, $t_d$ is the propagation time from Alice to Bob, $T$ is the time period of laser pulses, and $T_f$ is the temporal filtering window used.

Using Eq. \ref{eq:sifting}, the algorithm computes the parameter $n_i$, representing the clock cycle during which a specific quantum state was transmitted. In our implementation, the time period of the laser, denoted by 
$T$, is set to $200$ ns corresponding to $5$ MHz repetition rate. The optical path delay between Alice and Bob is approximately $666$ ns, which corresponds to a  200-meter channel. With all relevant parameters known, this allows Bob to accurately identify the transmitted state for each clock cycle. Consequently, events corresponding to incorrect basis measurements can be reliably filtered out, ensuring the validity of the sifted key. For the EPCD protocol, the sifting algorithm remains largely the same as in standard BB84, with the key difference being the need to log the time stamp of decoy pulses at Alice and record coincidence detection events at Bob.

\subsubsection{\textbf{Parameter Estimation}}

After key sifting, the raw keys of Alice and Bob may differ due to the QBER. To perform this estimation, a small portion of the sifted key is disclosed, typically around 10$\%$. The core idea is that this disclosed subset should be statistically representative of the entire key, assuming that errors are uniformly distributed across all bits. This uniform distribution ensures that even a small sample can accurately reflect the overall error rate. 
\begin{figure}[ht]
     \centering
     \begin{subfigure}[b]{0.42\textwidth}
        \centering
         \includegraphics[width=5.5cm]{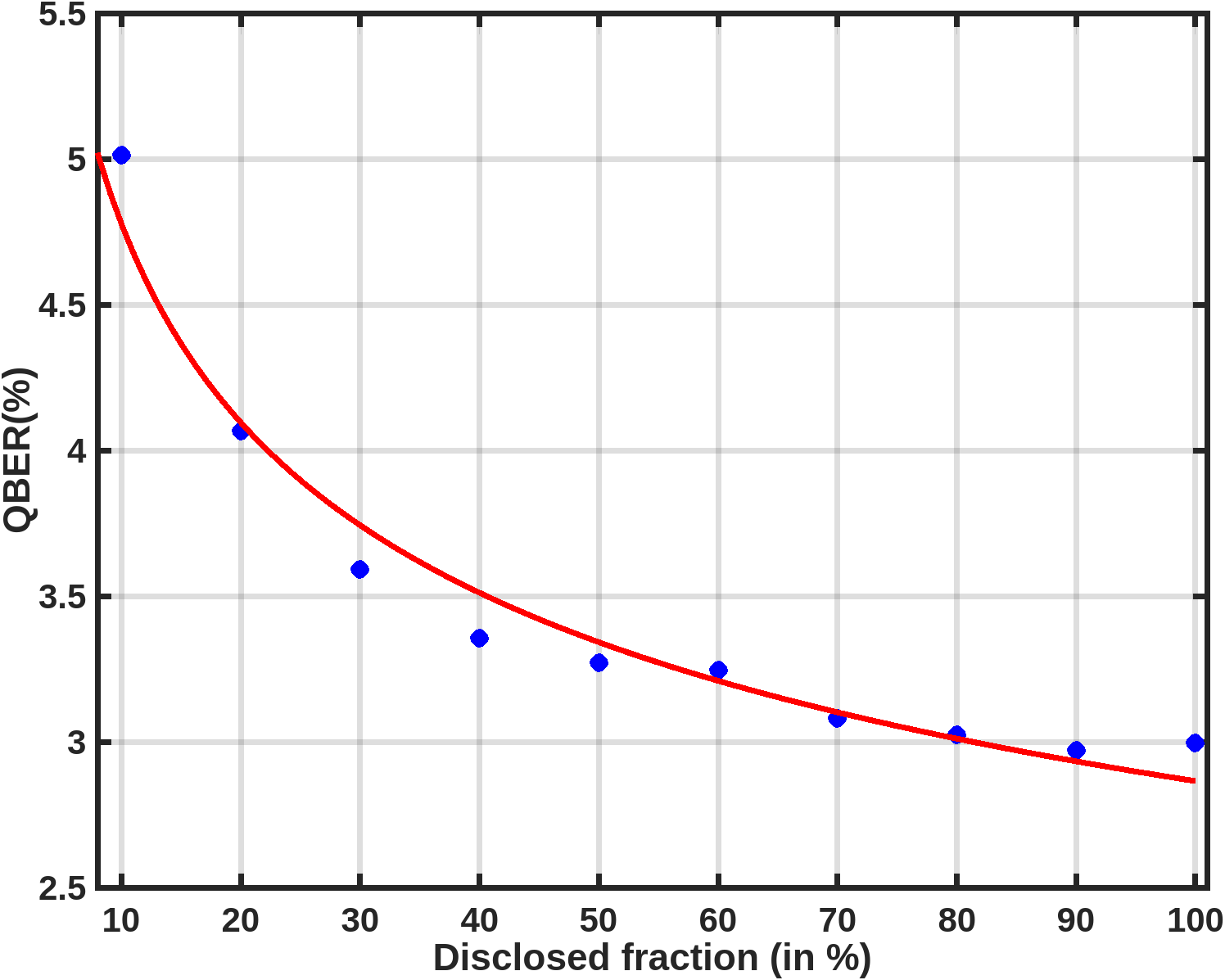}
         \caption{\footnotesize Graph shows the variation of QBER with the fraction of sift key bits disclosed. Fractions of total sift key bits starting from 10$\%$, 20$\%$,... 100$\%$ were disclosed, and QBER was calculated. The line is for the aid of the eye.}
         \label{fig:qberblock}
     \end{subfigure}
     \hspace{0.5cm}
     \begin{subfigure}[b]{0.42\textwidth}
         \centering
         \includegraphics[width=5.5cm]{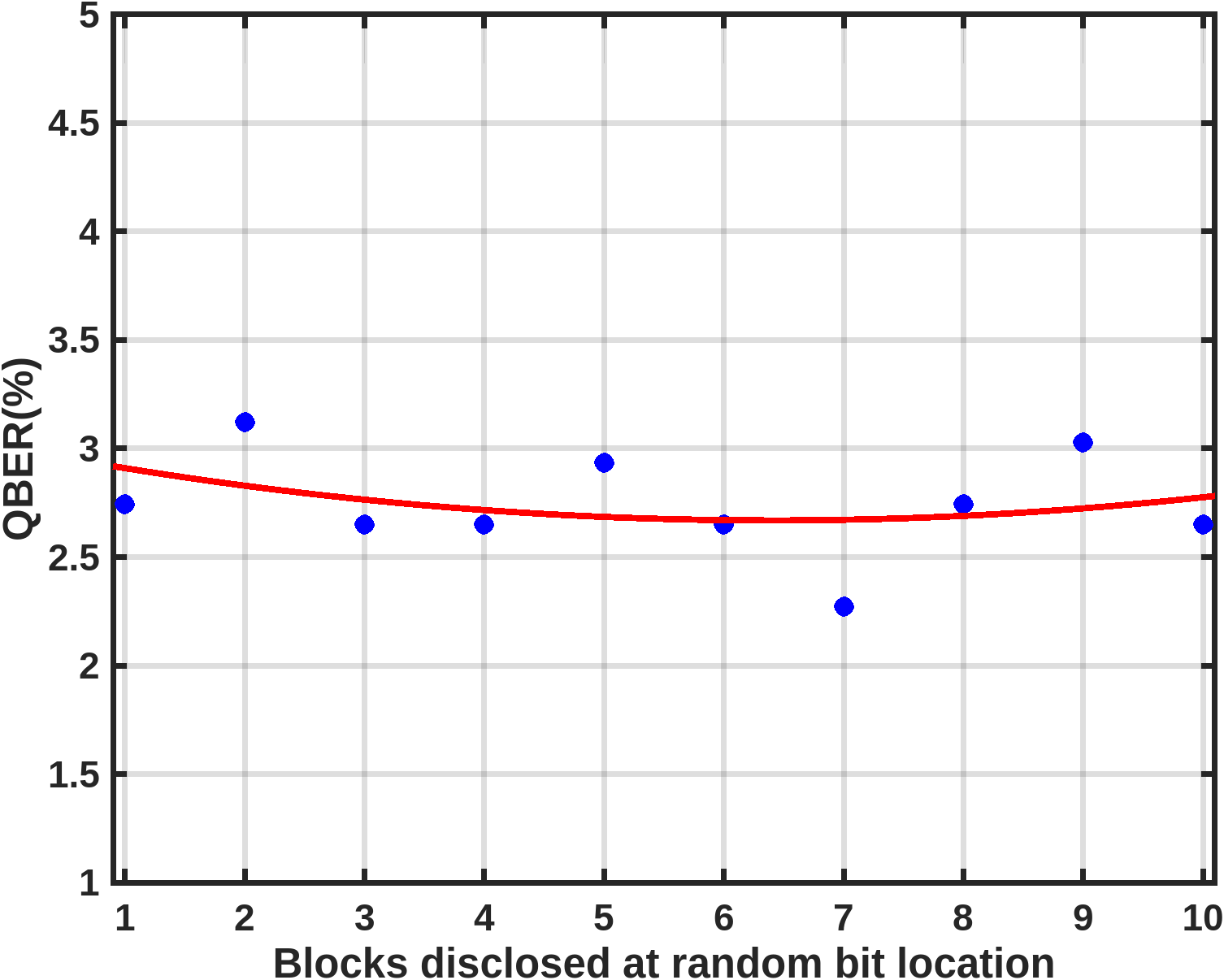}
         \caption{\footnotesize This graph shows the variation of QBER with disclosed fraction of sift key bits. A fixed fraction (10$\%$) of sift key bits is disclosed at the random bit location of sift key bits. The line drawn is for the aid of the eye.}
         \label{fig:qberrandom}
     \end{subfigure}
        \caption{Variations of QBER with disclosed fraction and block location of sift key}
        \label{fig:temp_wind}
\end{figure}
Our analysis, illustrated in Fig.\,\ref{fig:qberblock}, shows that while there may be slightly higher variations in QBER when only 10$\%$ or 20$\%$ of the sifted key is revealed, the estimated error rate stabilises and remains consistent with an increase in the sample size.
As the sample size increases, the accuracy of estimating the QBER improves due to reduced statistical fluctuations. A larger sample provides a more reliable approximation of the actual QBER, allowing for more precise and efficient error correction. Fig.\,\ref{fig:qberblock}    shows that beyond 30$\%$ sampling, the error in estimating the QBER decreases, making it an effective choice for parameter estimation. 

We also show in Fig.\,\ref{fig:qberrandom} that random sampling provides a reliable estimate of the true QBER. Since the errors are uniformly spread and the sample is chosen randomly, the process avoids bias and prevents any exploitable patterns, preserving both the accuracy of error estimation and the secrecy of the remaining key. The 2-3 $\%$ QBER observed in the experiment is intrinsic to the system and arises primarily from imperfections in the optical components such as the beam splitter (BS), polarising beam splitter (PBS), and half-wave plate (HWP). It represents the minimum achievable QBER in our setup, below which further reduction is not feasible because of these inherent optical limitations.

\begin{figure}
     \centering
     \begin{subfigure}[b]{0.45\textwidth}
        \centering
         \includegraphics[width=5.5cm]{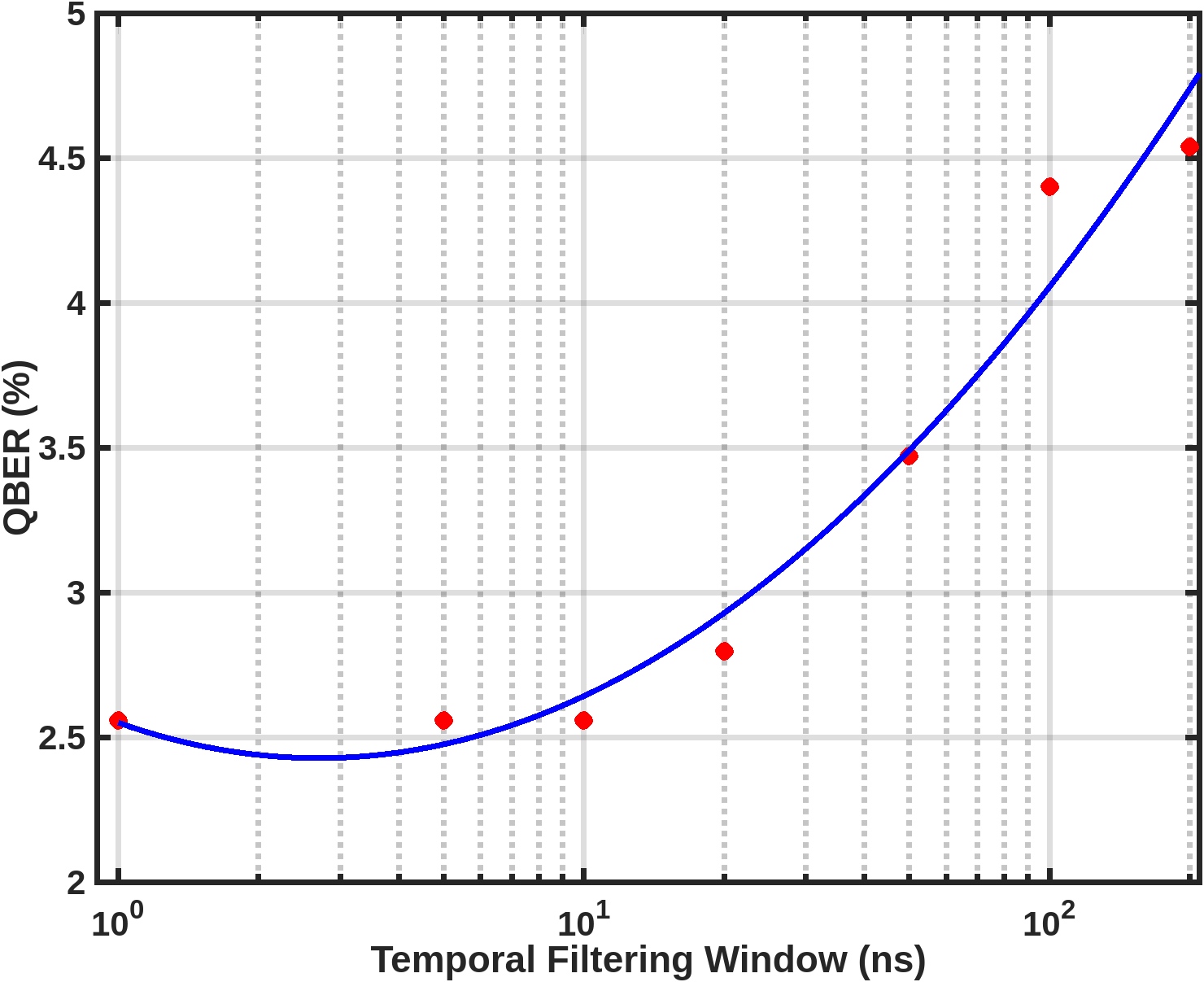}
         \caption{\footnotesize QBER ($\%$) as a function of temporal filtering window. The graph is plotted on a logarithmic scale for better visualisation. The line drawn is for the aid of the eye.}
         \label{fig:tempqber}
     \end{subfigure}
     \hspace{0.5cm}
     \begin{subfigure}[b]{0.45\textwidth}
         \centering
         \includegraphics[width=5.5cm]{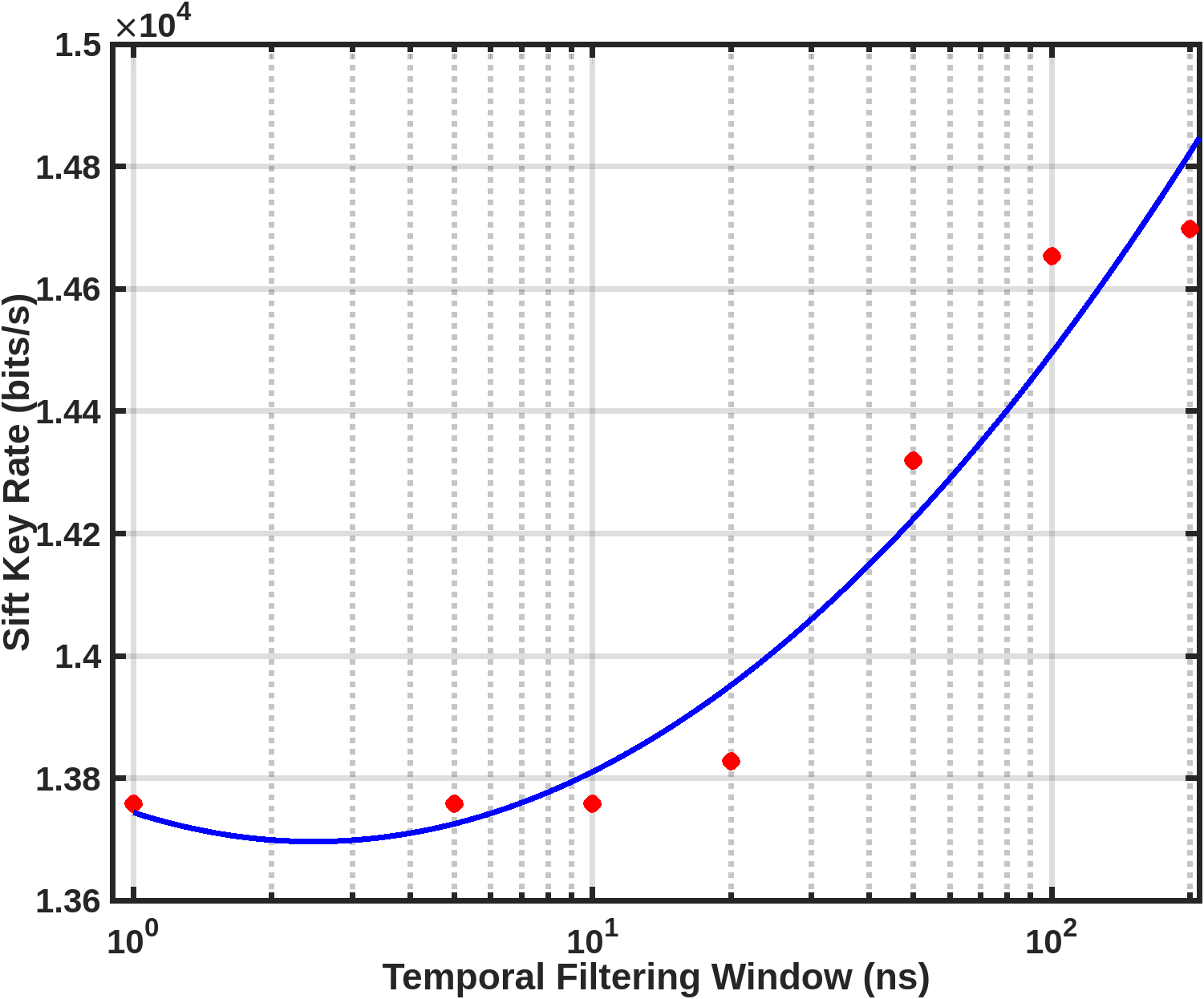}
         \caption{\footnotesize Key rate in bits/second as a function of temporal filtering window. The graph is plotted on a logarithmic scale for better visualisation. The line drawn is for the aid of the eye.}
         \label{fig:tempkeyrate}
     \end{subfigure}
        \caption{ Variation of QBER and key rate with temporal filtering window.}
\end{figure}

We further explore the effect of the temporal filtering window on the achieved key rate. Temporal filtering is employed to discard the detections arising from background light. We analyse the effect of the temporal window by varying the temporal window from 1ns to 200ns, where 200ns corresponds to the repetition period of the laser diodes. It is evident from Figs.\,\ref{fig:tempqber} and \ref{fig:tempkeyrate} that increasing the temporal filtering window increases the key rate, but also raises the corresponding QBER. This is due to the increasing amount of background photons which get matched during the sifting procedure. The FPGA used for timestamp recording operates at a clock rate of 100 MHz, which corresponds to a 10 ns time window. As a result, the QBER remains unchanged up to a 10 ns window, as shown in Figs.\,\ref{fig:tempqber},\ref{fig:tempkeyrate} and starts increasing after 10 ns.

\subsubsection{\textbf{Error correction}}

In classical communication, error correction codes are used to correct errors caused by channel noise. This is achieved by augmenting the message with additional bits, known as parity bits (r). To create error correction codes, a transmitter utilises a generator matrix (G), while a receiver uses a parity check matrix (H) to compute syndromes for decoding.

In QKD, decoding occurs at either Alice's or Bob's node, unlike classical communication, where both sender and receiver participate. Only syndromes are exchanged between Alice and Bob. Keys are adjusted based on these syndromes until both nodes possess identical keys \cite{mink2012ldpc}. 

The error correction process can be executed by considering either Alice's or Bob's bits as correct, as depicted in the schematic in Fig.\,\ref{fig:LDPC}. 
The following are the steps involved in the error-correction process:
\begin{figure}[h]
    \centering
     \includegraphics[width=13cm]{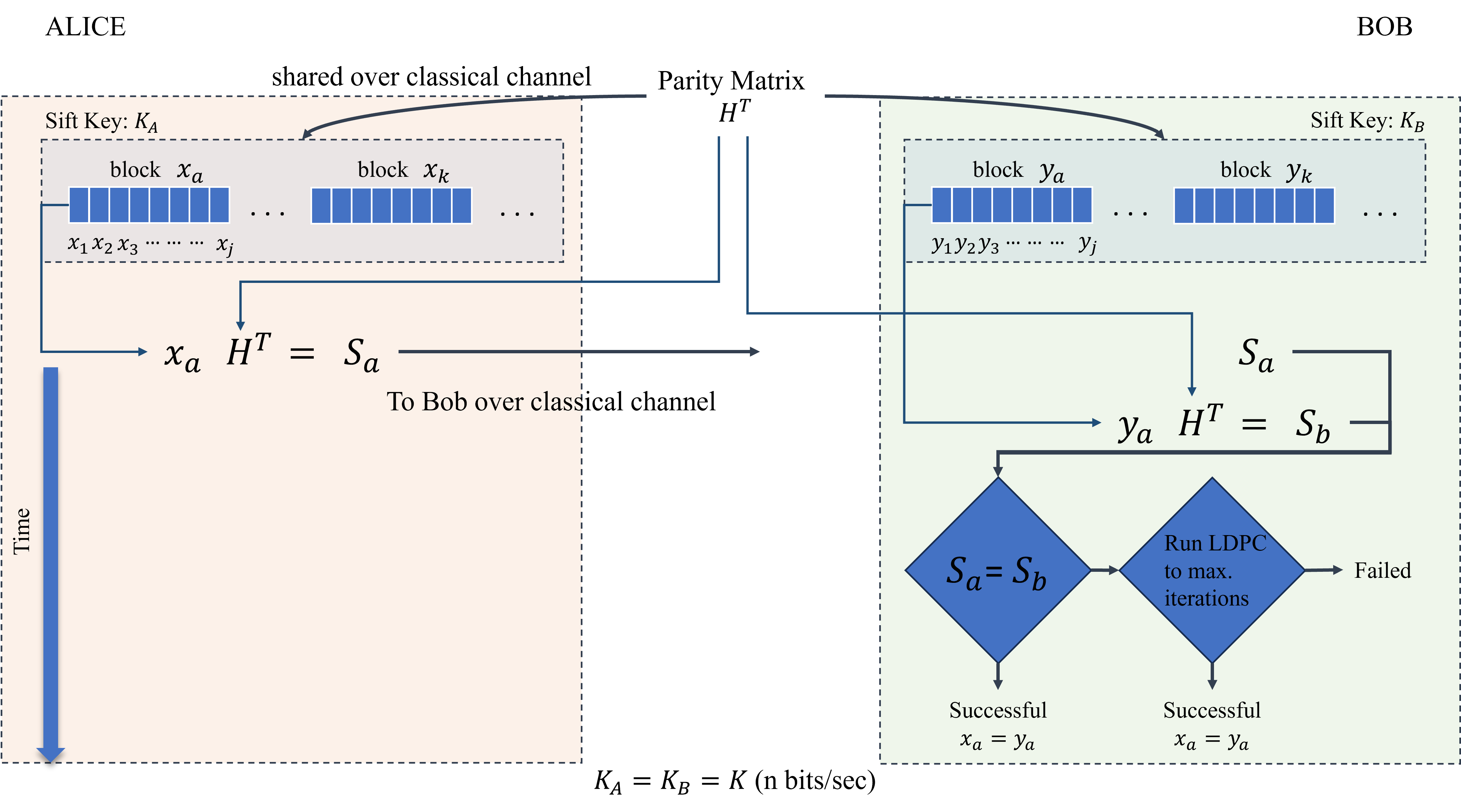}
    \caption{Flow diagram of error correction execution steps at LDPC decoder. }
    \label{fig:LDPC}
\end{figure}
Alice and Bob have sifted keys, $K_A$ and $K_B$, of equal length $n$, where the keys are not identical due to errors introduced in a few bit positions that need to be corrected. The sifted keys at Alice  
and Bob are partitioned into blocks ($x_a$, $y_a$) of length $j$, and low-density parity check (LDPC) codes are used for error correction, a technique extensively validated in classical communication systems.
Both Alice and Bob compute syndromes, $S_a$ and $S_b$, which are the products of the parity-check matrix and the sifted keys. Syndromes are exchanged and are then compared, and LDPC decoding is performed iteratively to correct the errors. If there are no errors in the sifted keys, \(S_a\) and \(S_b\) would be identical. If, after a maximum number of iterations, the syndromes still do not match, it indicates that the errors cannot be corrected, and the corresponding blocks are discarded.

In our error correction process, we have utilised LDPC coding with a block length of 1024 bits and a code rate of 1/2. Although the decoder is capable of performing up to 25 iterations,  in our case, only 5 to 7 iterations were generally sufficient to correct errors. Due to the relatively low QBER of 2-3$\%$, the error correction was highly effective, with almost zero probability of failure.

\subsubsection{\textbf{Privacy Amplification}}

In QKD, privacy amplification (PA) is a crucial step that ensures any partial information an eavesdropper (Eve) may have gained about the shared key becomes negligible. Eve's knowledge could arise from various sources, such as imperfections in the source (e.g., diode characteristic mismatch), interception of signals during transmission and the information disclosed during error correction.

To mitigate this, PA uses universal hash functions that compress the key in a way that effectively removes Eve’s potential knowledge about keys. The two-universality of the Toeplitz hashing method is particularly important, as it sets stringent conditions on how much useful information can be eliminated from Eve. Furthermore, the quality of the hashing method used to perform PA also depends on the seed used in privacy amplification \cite{tang2019high}.

The hashing method used in privacy amplification distils the secret key ($r$) from the raw key ($n$) generated through communication over the quantum channel. As the term distillation implies, this process compresses the longer raw key into a shorter secure key. This reduction enhances the security as the two-universality of the hashing function defines the one-way mapping of the raw key ($n$) to the secret key($r$). The relationship among the raw key ($n$)
secret key ($r$) and information potentially leaked to Eve, $t$ is defined by 
\begin{eqnarray}
r= n-t-s.
\label{PA_Eq}
\end{eqnarray}
The information gained by Eve, $t$, can be further decomposed into several contributing factors, depending on the experimental implementation. $t$ primarily comprises information obtained during transmission through the quantum channel, which is QBER ($e$) and the parity bits ($p$) revealed during the error correction process. Thus, the relationship becomes,
\begin{eqnarray}
t=2 \ n \ e+p.
\end{eqnarray}
The security parameter $s$ puts a bound on the mutual information between Alice and Eve, $I(A:E)$, representing the maximum amount of information leakage that Alice and Bob can tolerate while still considering the key secure. The bound is defined as,  
\begin{eqnarray}
I(A:E) \leq \frac{2^{-s}}{ln 2}.
\end{eqnarray}
\noindent
The value of $s$ varies from one experimental realisation to another in the QKD protocol. 

In our experiments, the mutual information between Alice and Eve is around 0.001, which corresponds to a security parameter $s$ of 10. This implies that Alice and Bob can securely perform QKD with $s$ =10 as long as the mutual information remains within this range.

Furthermore, the seed used in privacy amplification is typically taken from a random number generator with good uniformity features over the range ($0,1$). The Toeplitz matrix used for privacy amplification is built using this seed. The length of the seed is typically defined as $n + r-1$. Therefore, the seed length used to generate the Toeplitz matrix, which determines the extent to which Eve’s information is effectively eliminated, depends on the experimental realisation. 
The expression for the matrix derived from the seed bits is
\begin{eqnarray}
    \label{eq:top_PA}
    T_{ij}=b_{r-1+j-i},
\end{eqnarray}
where $T_{ij}$ is the element of the Toeplitz matrix, index $i$ will run up to $r$ bits and the maximum value of $j$ will be $n$ bits. Multiplying the Toeplitz matrix with the sift key will give the secret key of $r$ bits
\begin{eqnarray}
    \label{eq:top_PA2}
    [r]_{r \times 1} = [ T]_{r \times n} [n]_{n \times 1}.
\end{eqnarray}
Finally, the quality of universality of the hashing method defines the quality of one-way mapping. The 2-universality of the hash function, a stricter subset of 1-universal hash functions, means that when considering inputs $n_1$ and $n_2$ and their corresponding outputs $r_1$ and $r_2$ in a hash function, the probability of collision between ($n_1$, $r_1$) and ($n_2$, $r_2$) is less than $1/d^2$ where $d$ is the set of values spanned by $r$. The probability of collision ensures that the hashing process is impartial.

\section{Results and discussion}\label{sec:RnD}

\begin{figure*}[htp]
    \centering
\includegraphics[scale=0.18]{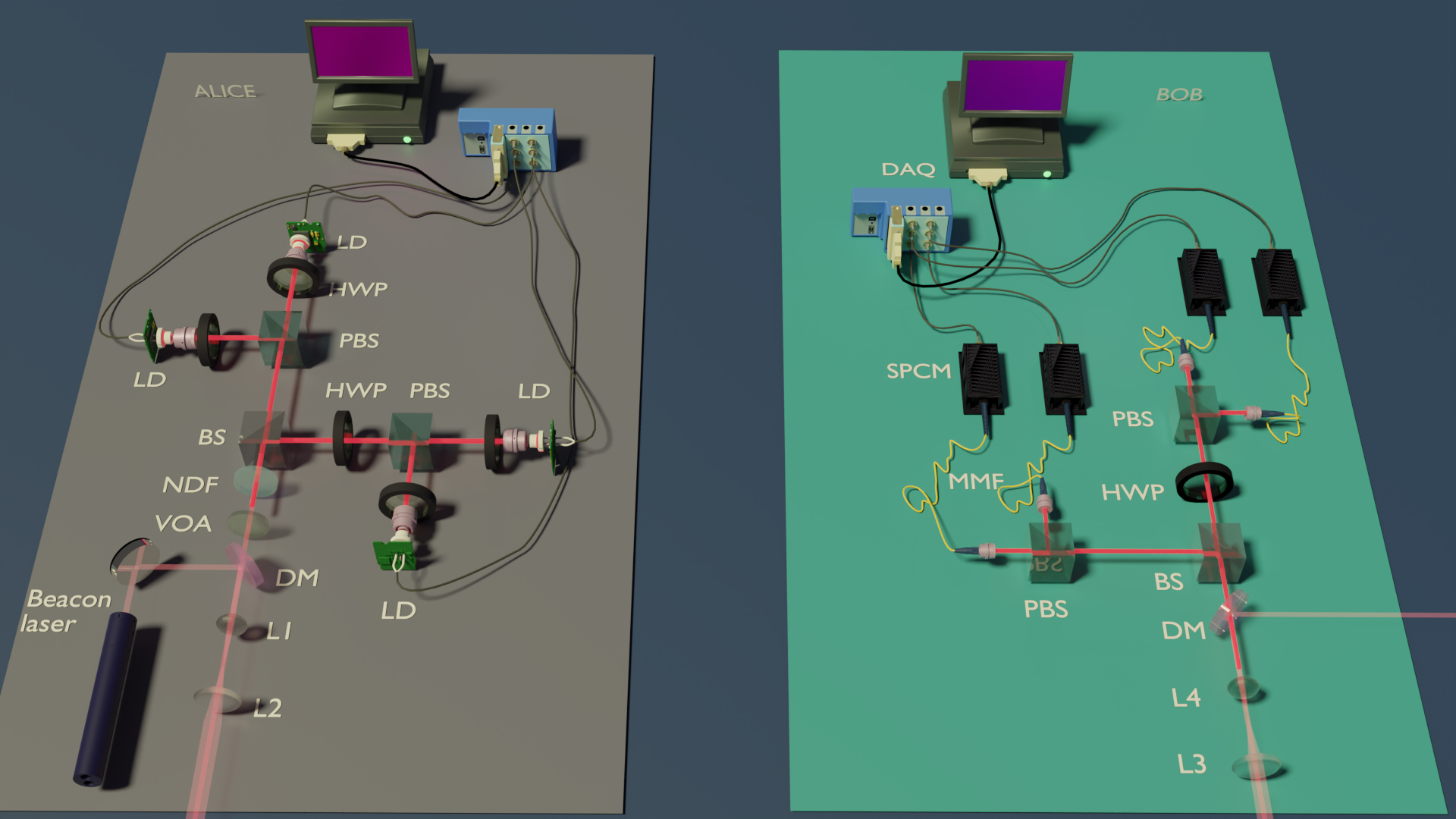}
    \caption{Schematic of the experimental setup of BB84 with EPCD protocol. It includes both the optics and electronic components. LD: Laser Diodes, HWP: Half Wave Plate, BS: Beam Splitter, PBS: Polarising Beam Splitter, NDF: Neutral Density Filter, VOA: Variable optical Attenuator, DM: Dichroic Mirror, L: lens, MMF: Multi-Mode Fibre, SPCM: Single Photon Counting Module, DAQ: Data Acquisition system.}
    \label{fig:setup}
\end{figure*}

In this section, we present the experimental results of the QKD protocol implemented over a 200-meter free-space communication link.
The BB84 QKD protocol was successfully implemented for a 200-meter free-space channel. A detailed schematic of our implementation setup is shown in Fig.\,\ref{fig:setup}. 
System components (source, channel, detectors) were thoroughly characterised before performing the experiment. Since the channel length in our implementation is relatively short (200 meters), a dedicated pointing and tracking system was not required. Instead, the optical alignment between the transmitter and receiver was manually optimised.
Once alignment was optimised, photons were transmitted through the free-space channel, with launching optics directing the quantum signal towards Bob. The QKD protocol execution steps were automated using LabVIEW FPGA software. 
Once synchronisation is established for the free space channel, protocol execution starts, which includes enabling lasers for state preparation, recording the time stamps of photons at Alice and Bob, and execution of the TCP/IP protocol for further post-processing. In our implementation, we present a very simple and efficient sifting algorithm which takes very little time for basis sifting. We also show in Fig.\,\ref{fig:qberrandom} that random sampling of sift key bits for error estimation gives better QBER estimation and the same QBER through all sifted key bits. This shows that in our implementation, the error was uniformly distributed throughout the sifted bits. In our experiment, the QBER is consistently observed to be in the range of 2–3 $\%$.

In this work, we also describe how to choose the optimal detection window for valid photon detection. We have thoroughly analysed the effect of the temporal window on QBER and key rate. We define a valid detection window centered on the expected detection time, within which detections are considered valid. The window size is a trade-off; a larger window improves detection chances but raises the QBER, while a smaller window lowers errors but decreases the key rate by rejecting more valid detections. Proper selection of the detection window balances the QBER and key rate, optimising QKD performance. Our results mentioned in Figs.\,\ref{fig:tempqber} - \ref{fig:tempkeyrate} show that we can keep our detection window up to 10 ns. If we increase the window to more than 10 ns, then QBER starts increasing, which is not desirable. The increase in sift key rate follows the QBER, as we have more keys with errors.
\begin{table}[h]
    \centering
    \begin{tabular}{|c|c|}
    \hline
\textbf{Parameter} & \textbf{Values}\\ \hline
 FPGA clock rate @ Alice & 100 MHz \\ \hline
  FPGA clock rate @ Bob & 100 MHz \\ \hline
Laser repetition rate& 5 MHz  \\ \hline
Laser optical pulse width & $1$ ns \\
\hline
    Mean photon number  (\(\mu\)) & 0.14 \\ \hline
  Detection efficiency ($\eta_{d}$)& 62 $\% $ \\ \hline
Detector dark counts & $65$ cps \\ \hline
 Detection jitter & $350$ ps \\
   \hline
 Coupling efficiency ($\eta_{c}$)& $79\%$ \\ \hline
  Channel transmission($\eta_{ch}$) & 80 $\% $  \\ 
\hline
QBER & 2-3$\% $ \\ 
\hline
Sift key rate  & 137 kbps \\ \hline
Secure key rate  & 61 kbps \\ 
\hline
\end{tabular}
\caption{QKD system parameters, dark counts of detectors are mentioned in cps: count per second. The sift key rate and secure key rate are mentioned in kbps: kilobits per second. } 
 \label{table: system}
\end{table}

In Table\,\ref{table: system}, we show the key parameters that require attention while implementing the QKD protocol in real time.

\begin{figure}[h]
    \centering
\includegraphics[width=8cm]{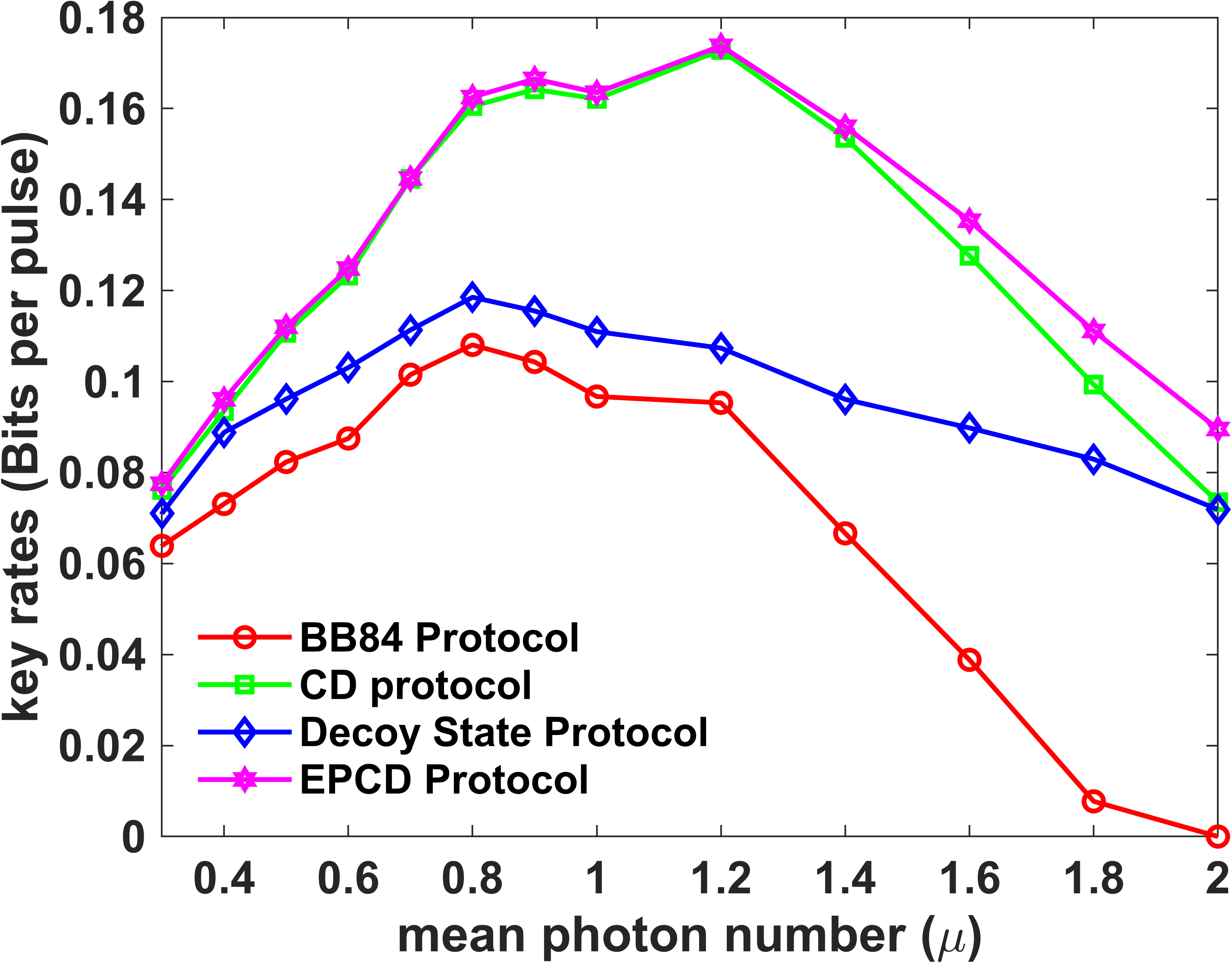}
\caption{Experimental key rates (bits per pulse) as a function of mean photon number ($\mu$) for four protocols: (i) BB84 Protocol, (ii) CD protocol, (iii) Decoy state protocol and (iv) EPCD protocol.}
    \label{fig:keyvsmu}
\end{figure}
In this work, we have extended our implementation to include the Entrapped Pulse Coincidence Detection (EPCD) protocol over a 200-meter free-space channel. By combining the EPCD scheme with the standard BB84 framework, we harness the advantages of both approaches. In our implementation, entrapped pulses with mean photon number $\nu \approx 0.1$ are used, while the signal intensity is varied from $\mu = 0.1$ to $\mu = 2$ to determine the optimal value. Coincidence monitoring for signal and entrapped pulses enhances the system's ability to detect realistic photon-number splitting (PNS) attacks. Under conditions where no PNS activity is observed, incorporating two-photon events contributes to a higher secure key rate shown in Fig.\,\ref{fig:keyvsmu}.
Our results highlight the feasibility of free-space QKD  in real-world conditions.

\section{ \label{sec:conc}Summary and Conclusion}

This work presents a practical implementation of the QKD protocol in a real-world setting. We have provided a detailed overview of the necessary steps and highlighted the crucial details required during the protocol's field deployment. Additionally, we have discussed the important system parameters, such as the key rate and QBER, which are influenced by factors like the laser diode driving speed, the quality of clock synchronisation, the temporal filtering window and the effectiveness of the post-processing scheme. By addressing these elements, we have demonstrated how to effectively carry out the QKD protocol in a real-world scenario. This work can be useful and act as a manual for the field implementation of QKD protocols. Our study also demonstrates the significant advantages of our proposed EPCD protocol over the standard BB84 and Decoy State protocols. This comparison highlights the EPCD protocol's capability to achieve higher key rates, especially over longer distances. These findings provide a promising pathway to attain enhanced key rates for extended communication distances.

\section*{Acknowledgement}

The authors thank the group members of the QST laboratory for their valuable input and support. The authors acknowledge funding from the Department of Space (DoS) and partial financial support from DST through the QuST program. Special thanks to Dr. Satyajeet Patil and Dr. Vardaan Mongia for their assistance in optical alignment and to Mr. JayaKrishna Meka for his technical support. PC thanks  Dr. Prashant Kumar for his initial guidance on the sifting algorithm and Mr. Chandan Kumar for his valuable support in PCB designing for the laser driver circuit.

%%%%%%%%%%%%%%%%%%%%%%%%%%%%%%%%%%%%%%%%%%%%%%%%%%%%%%%%%
%                    ~Disclosure~                       %
%%%%%%%%%%%%%%%%%%%%%%%%%%%%%%%%%%%%%%%%%%%%%%%%%%%%%%%%%
\section*{Disclosure}
The authors declare no conflicts of interest.

%%%%%%%%%%%%%%%%%%%%%%%%%%%%%%%%%%%%%%%%%%%%%%%%%%%%%%%%%%
%                    Appendix                            %
%%%%%%%%%%%%%%%%%%%%%%%%%%%%%%%%%%%%%%%%%%%%%%%%%%%%%%%%%%

%\bibliographystyle{plain}
\bibliography{main}

\end{document}